\documentclass[lettersize,journal]{IEEEtran}
\usepackage{amsmath,amsfonts}
\usepackage{algorithmic}
\usepackage{algorithm}
\usepackage{array}
\usepackage[caption=false,font=normalsize,labelfont=sf,textfont=sf]{subfig}
\usepackage{textcomp}
\usepackage{stfloats}
\usepackage{url}
\usepackage{verbatim}
\usepackage{graphicx}
\usepackage{cite}
\usepackage{xcolor}
\usepackage{hyperref}

\begin{document}

\title{Dimensionality reduction techniques to support insider trading detection}

\author{Adele Ravagnani, Fabrizio Lillo, Paola Deriu, Piero Mazzarisi, Francesca Medda, Antonio Russo \thanks{Adele Ravagnani is with Scuola Normale Superiore, Pisa, Italy.}\thanks{Fabrizio Lillo is with Scuola Normale Superiore, Pisa and Dipartimento di Matematica, Universit\`a di Bologna, Italy.}\thanks{Paola Deriu and Antonio Russo are with Commissione Nazionale per le Società e la Borsa (Consob), Italy.}\thanks{Piero Mazzarisi is with Dipartimento di Economia Politica e Statistica, Universit\`a di Siena.} \thanks{Francesca Medda is with University College London, London, UK and Commissione Nazionale per le Società e la Borsa (Consob), Italy.}\thanks{This work has been submitted to the IEEE for possible publication. Copyright may be transferred without notice, after which this version may no longer be accessible.}}

\maketitle

\begin{abstract}
Identification of market abuse is an extremely complicated activity that requires the analysis of large and complex datasets. We propose an unsupervised machine learning method for contextual anomaly detection, which allows to support market surveillance aimed at identifying potential insider trading activities. This method lies in the reconstruction-based paradigm and employs principal component analysis and autoencoders as dimensionality reduction techniques. The only input of this method is the trading position of each investor active on the asset for which we have a price sensitive event (PSE). After determining reconstruction errors related to the trading profiles, several conditions are imposed in order to identify investors whose behavior could be suspicious of insider trading related to the PSE. As a case study, we apply our method to investor resolved data of Italian stocks around takeover bids. 
\end{abstract}


\section{Introduction}
\IEEEPARstart{I}{nsider} trading is the unlawful practice of trading by exploiting nonpublic confidential information about a listed company. It is a type of market abuse: it prevents full and effective market integrity, it violates natural demand-supply dynamics, it compromises public confidence. The announcement of confidential information to the market i.e. price sensitive events (PSEs), implies a specific price response. For instance, after a PSE, as the announcement of a takeover bid, the price is likely to increase. Therefore, foreknowing a PSE can be easily exploited to make a profit. 

Most jurisdictions around the world prohibit or criminalize such a type of practice \cite{bhattacharya2002}. However, each country has specific rules and enforcement. In the European Union, insider trading is expressly
prohibited and administratively sanctioned (Directive 2014/57/EU so-called MAD II, European Regulation 596/2014 so-called MAR) \cite{eu_reg}. In Italy the first Market Abuse Directive (MAD) passed in 2005, thereby introducing stringent administrative sanctions in addition to the pre-existing criminal penalties \cite{consob_regulation}.

However, the ``proof'' and the subsequent imposition of a sanction (either administrative or criminal) to an investor that has operated as an {\it insider} is a complex process, involving many steps: (i) the detection of alerts pointing to anomalous behaviors, (ii) the concrete assessment of the allegedly suspicious conduct, (iii) the investigation phase aimed at gathering evidence and clues of the abusive conduct, and (iv) the subsequent legal trial to confirm the unlawful conduct.

As part of an ongoing collaboration between academia and the Italian market supervising authority (Consob),  in \cite{our_paper} we recently proposed a methodology which, focusing on the first and partially second step above, supports the monitoring and surveillance processes by the competent Authority and the assessment of the trading conduct. This is done by focusing on the discontinuities in the trading activity of single investors with respect both to their normal activity and to the behavior of their peers, i.e. investors with similar activity. The peers are identified by using an unsupervised dynamic clustering approach based on k-means which takes as input three trading features (termed signed turnover, magnitudo, maximum exposure) obtained from the trading time series of each investor. The choice of the features was motivated by their financial meaning and by explanability reasons, since one can easily interpret which features are responsible for the anomalous behavior.

However, choosing to represent a long time series with three scalar features is somewhat arbitrary and might hide important information useful to identify anomalies in the trading activity of an investor. For this reason, in this paper we present a new approach aiming to overcome this limitation by employing dimensionality reduction methods, namely Principal Component Analysis and Autoencoders. This approach maps the whole time series in a set of coordinates in a suitable latent space which retains a large part of its variability. Large deviations between the original time series and the one reconstructed from the latent space indicate potential anomalies, following the so-called {\it reconstruction-based} paradigm.
In summary, the idea is that the model should learn by itself the features that are more relevant to investors’ characterization. The methodology is tested on a unique dataset, provided by Consob and containing the trading activity of each investor in the Italian stock market in the proximity of several PSEs.

\subsection{Related literature}
\noindent Our work suits into the framework of anomaly detection. This field has been widely explored in the literature, especially in recent years, when its developments have been moving at the same pace as machine learning. The applications in the field of financial fraud detection are numerous \cite{west2016} and, among them, several works are related to the detection of market abuse such as \cite{ahern, augustin, baltakiene2022, donoho2004, frino, golmohammadi2014,  keown, li2017, lamorgia2022, meulbroek, minenna2003, park, thalassinos}. In particular, \cite{ahern, augustin, baltakiene2022, donoho2004, frino,  keown, meulbroek, park, thalassinos} investigate insider trading. By employing supervised approaches, they focus on the identification of market conditions prone to insider trading and on the characterization of insiders' trading behaviors (for more details, refer to the papers and to \cite{our_paper} for a brief summary of them). Our previous work \cite{our_paper} as well as this new paper, differ from previous literature because of our goal and unavailability of labels. The latter force us to rely on an unsupervised approach. Concerning the goal, we aim to identify investors engaging in suspicious activities related to PSEs. This task is feasible thanks to the transaction reporting database provided to us. This dataset comprehensively covers the daily activity of all investors trading Italian stocks over a period of approximately three years, enabling us to effectively track the trading activity of each individual investor.

Anomaly detection amounts at identifying data instances that deviate from normal behavior and that are uncommon within the data set. Its aim is to identify a region of the features' space where normal observations lie; observations that do not belong to this region are defined as anomalies \cite{AD_review}. Determining this normality region is not straightforward and is accompanied by several difficulties. The formulation of an anomaly detection method involves four main aspects \cite{AD_review}: 
\begin{enumerate}
    \item the availability of data labels causes the employment of a different approach: supervised when each observation is labeled as normal or anomalous, semi-supervised when training data do not contain any anomalies and unsupervised when no labels are provided as in our interest case;
    \item the desired output of the technique is usually a score, which is associated with each observation and quantifies the magnitude of its anomalous character. Setting a suited threshold, each data instance is labelled;
    \item input data can be of various type (binary, categorical, continuous, univariate, multivariate) and independent among them or related to each other. Time series and sequences, spatial data and graph data are examples of dependent data and ad hoc methodologies have to be employed for them \cite{AD_graph_review, aggarwal2013};
    \item anomalies can manifest in different forms. \textit{Point anomalies} are single elements and they could be \textit{global} or \textit{local} depending on whether the entire feature space or a specific region of it is considered. {\it Contextual anomalies}, also termed {\it conditional anomalies} \cite{song2007} require that a given context is taken into account in order to identify them. Finally, \textit{collective anomalies} have data instances that are anomalous if they are considered together and they can occur in a data set where data instances are dependent. 
\end{enumerate}

It is evident that anomaly detection problems are challenging, especially in the unsupervised setting. A variety of different approaches have been developed to address them. In particular, the main paradigms in time-series are: \textit{clustering-based, distance-based, reconstruction-based} and \textit{forecasting-based} \cite{mejri2022}. Among them, the methods are multiple and their formulations are case-by-case dependent.

In this work, we develop an approach which aims at identifying \textit{contextual anomalies} by relying on continuous and multivariate
data instances and lies in the \textit{reconstruction-based} paradigm of unsupervised anomaly detection \cite{mejri2022}. This framework aims at training models that reconstruct normal data instances well. 
In this way, we expect that anomalous data will be reconstructed with a large error. An anomaly is detected when the reconstruction error is greater than a threshold i.e.
\begin{equation*}
    ||X - \hat{X}|| > \delta
\end{equation*}
where $X$ is an original data sample, $\hat{X}$ is its reconstructed counterpart and $\delta$ a suitable threshold. Models' performances are usually compared in terms of the most common metrics e.g. precision, recall and F1-scores \cite{mejri2022}. 

Our case is even more complicated since we are not provided with labels which allow to compute the metrics to assess the models' performances. Therefore, in order to compare the results of different reconstruction models, we have to rely on qualitative inspections.

The standard model employed in \textit{reconstruction-based} approaches is Principal Component Analysis (PCA). Its goal is to obtain a compressed representation of data, retaining the most important features. Data are mapped to a lower dimensional space by orthogonal transformations that aim at maximizing data variance or equivalently, minimizing reconstruction error \cite{pca_book}. Also variants of this method have been employed in the anomaly detection literature e.g. \cite{lee2013}.

The nonlinear counterpart of PCA is an autoencoder \cite{goodfellow}. As PCA, autoencoders's goal is to minimize reconstruction errors but, in this case, the compression and decompression steps are made by means of neural network layers. Multiple are the approaches proposed in the anomaly detection literature which rely on autoencoders e.g. \cite{hojjati2023}. Complex autoencoder architectures can be devised, as deep, convolutional, LSTM, variational autoencoders \cite{zhang2023}. Also, generative adversarial networks have been employed \cite{audibert, liu2023} as well as combined approaches \cite{crepey2022}.

If an autoencoder is provided with one hidden layer and linear activation functions, the analogy with PCA is evident and in the literature, it has been investigated in several works. In particular, in \cite{kunin2019}, the authors characterize the loss landscapes of linear autoencoders (LAEs), prove that LAEs with $L_2$ regularization learn the PCA's principal directions and provide an algorithm to recover them from LAEs' results. 

{\bf Contributions of the paper and outline}\footnote{The methodology presented in the paper was conceived in 2023 for the purpose of developing a proof of concept. It is, in no way, a tool used in the analysis and investigations carried out by Consob. The methodology may possibly constitute the future one of the tools to help and support the preliminary analysis and detection activities more efficiently. Any subsequent enforcement activity will, in any case, be based on the broader set of information that is gathered in the course of investigations and other possible types of analysis.}
The main contributions of the paper can be summarized as follows:
\begin{itemize}
\item We devise a method to support decision in insider trading detection which is not based on the definition of specific trading features;
    \item Our method is an unsupervised approach for contextual anomaly detection, without any labels to check results and compare performances;
    \item We apply principal component analysis and autoencoders for the reconstruction of trading profiles.
\end{itemize} 

The paper is organized as follows. In Section \ref{section_method}, the proposed method is described. Section \ref{section_data} presents the data set we use in our empirical analysis
and Section \ref{section_results} presents the results obtained by our method, with a special focus on
one PSE. Finally, conclusions are drawn in Section \ref{section_conclusion}. In the Appendices, some figures, which are explained in the main text, are reported, and other collateral issues are investigated.

\section{Method}\label{section_method}
\subsection{Overview}\label{subsection_method_overview}
\noindent As in \cite{our_paper}, we are tackling an unsupervised problem without any availability of labels and we consider a specific class of price sensitive events (PSEs), namely announcements of takeover bids. A takeover bid is a public offer made by a physical person or a legal entity who is willing to buy other shareholders' shares at a price higher than the stock market value. If investors know in advance when the announcement of the takeover bid will occur, they can exploit their information by buying before the PSE. Indeed, when the takeover bid occurs, the shares' price goes up aligning with the offer price and thus, the informed investors can sell by making a no-risk profit.

We focus on a single asset, the one for which we have a PSE, and on a time window with $T$ trading days. The first part of this time window - e.g. $6$ months - is a reference period and the second part - e.g. $1$ month - is an investigation period i.e. a short time window preceding the PSE that will be defined as $\Delta$ in the following. As a first step, we compute the trading position of each investor on each day. Given $N_0$ investors and $T$ trading days $\{t_0, t_1, \ldots, t_{T}\}$, the position of investor $i$ on day $t$ is defined as follows:
\begin{equation}\label{formula_position}
    x_i(t) = \sum_{t_0 \leq t' \leq t} [V_b(i,t') - V_s(i,t')] 
\end{equation}
where $V_{b/s}(i,t)$ is the number of shares bought/sold by investor $i$ on day $t$. Therefore, a vector $x_i = \big[x_i(t_0), \ldots, x_i(t_T)\big] \in \mathbb{R}^{T}$ is assigned to each investor. As usually done, we normalize data as
\begin{equation}\label{eq_norm}
    x_{i}(t) \rightarrow \frac{x_{i}(t)}{\max_t{|x_{i}|}} = \frac{x_{i}(t)}{||x_i||_\infty}
\end{equation} and investors with constant positions i.e. investors who do not trade or are strict daily investors (i.e. the number of shares purchased and sold on each day are equal), are discarded.

In the definition of Equation \ref{formula_position}, we assume that investors' positions are null on $t_0$. Of course, this is not true in general. However, since information on the precise composition of the portfolio of each investor is not available, this sounds as the best proxy of asset positions. In the following, we will see that actually this is not an issue for this new method, contrary to what happens for our previous method \cite{our_paper}.

We also observe that positions are computed using the number of shares and not Euro. The reason is that the monetary value of a portfolio fluctuates in response to the changing price and these fluctuations affect in the same direction positions with the same sign (e.g. long or short). Thus, spurious correlations between positions might be detected when using Euro as a unit of measurement. 

Indicating with $N$ the number of investors with non-constant positions, we end up with a data set $X \in \mathbb{R}^{N, T}$ with rows $x_i$. This data set is the input of a dimensionality reduction method (we are going to use PCA and autoencoders), which will allow to obtain a reconstructed representation of the data after a compression i.e. 
\begin{equation*}
    X \rightarrow  Z = f_1(X)\rightarrow \hat{X}= f_2(Z)
\end{equation*}
where $X, \hat{X} \in \mathbb{R}^{N,T}, \ Z\in \mathbb{R}^{N,K}$ and such that the reconstruction error is minimized with respect to the transformations $f_1$ and $f_2$ i.e.
\begin{equation*}
    \hat{X} = \arg\min_{\hat{X}'} ||X - \hat{X}' ||_F^2
\end{equation*}
where $|| \cdot ||_F$ is the  Frobenius norm.

In the compression phase, observations are mapped to a lower dimensional space that captures common and
essential characteristics. In our setting, the features which are subjected to compression are the positions of investors' on each day. This consists in identifying a subset of days or combinations of them with a major role in the characterization of agents' trading activity.

After the dimensionality reduction step, investors with anomalous activity (potential insiders) are identified following the \textit{reconstruction-based} paradigm and assuming they are substantially less numerous than investors with normal behavior. The preparatory steps of the method we develop are the following:
\begin{itemize}
\item compute the reconstruction errors
    \begin{equation*}\begin{split}
        \epsilon_i(t) &= |x_i(t) - \hat{x_i}(t)| \ \ i=1,\ldots,N, \ t=1,\ldots,T,\\
    \end{split}\end{equation*}
    we expect that normal observations have low $\epsilon_i$, while anomalies have large $\epsilon_i$;
    \item compute the anomaly scores i.e. the largest reconstruction errors for each investor:
\begin{equation*}
    s_i^* = \max_t \epsilon_i(t)  \ \ i = 1, \ldots, N; %
\end{equation*}
\item localize the largest reconstruction errors:
\begin{equation*}
    t_i^* = \arg\max_t \epsilon_i(t), \ \ i = 1, \ldots, N
\end{equation*}
which are such that
\begin{equation*}
    \epsilon_i(t_i^*) = s_i^*, \ \ i = 1, \ldots, N ;
\end{equation*}
    \item compute $n_{t} = \text{card}\{i: t_i^* = t\} \ \forall t = 1, \ldots, T$, where $n_t$ is the number of investors having the largest reconstruction error on day $t$;
    \item compute $d_i$ for $i=1, \ldots, N$ i.e. the number of activity days of each investor.
    \end{itemize}
Finally, in order to detect potential insiders, we devise a method which is based on the following idea. In order to be anomalous, an investor should satisfy the following conditions: (a) to have at least one day for which the reconstruction error is large, (b) the corresponding time lies in the investigation period, (c) she has either a small number of activity days or her identified anomalous score is on a day when not too many other investors do, and (d) she is in a net buying position on the day of the PSE. This last condition can be set by imposing that the difference between the position on the PSE and the position on the first day of the reference period is larger than a threshold, that we choose equal to $0.5$. 

Formalizing the above conditions, we say that an investor $i$ is anomalous if
\begin{equation}\label{eq_anom}\begin{split}
    \Bigg\{\Big( & \epsilon_i(t), t, n_{t}\Big):
      \epsilon_i(t) \geq \epsilon_{\theta}, \ t \in \Delta,\\
    & n_{t} < n_\theta \text{ if } d_i > d_{\theta} \text{ or } \forall n_t \text{ if } d_i \leq d_{\theta}, \\ & i  \text{ has a net buying position on the PSE} \Bigg\} \neq \emptyset. 
    \end{split}
\end{equation}

This criterion depends on three threshold parameters $d_\theta$, $\epsilon_{\theta}$, and $n_\theta$.
The parameter $d_\theta$, the minimal number of days in item (c) of the criterion above, is set to $3$. Instead, we choose $\epsilon_{\theta}$ and $n_\theta$ in a data driven fashion: we estimate the probability density function  of the anomaly scores $s^*_i$ and of the times of the largest reconstruction errors $t^*_i$. 
Since we observe that the former distribution is bimodal (see the left panel of Figure \ref{anomaly_scores_pdf}), we expect that normal investor profiles are associated to small anomaly scores $s_i^*$, while anomalous ones to high scores. Thus $\epsilon_{\theta}$ is chosen as the local minimum between the two modes of the distribution. In practice, we relied on the module \textit{signal} of the Python library \textit{scipy}. Finally, $n_\theta$ is chosen as the top decile of its distribution. 

Supervising authorities are often interested in a ranking of potential insiders to identify the most suspicious investors. Our approach is able to deliver such a ranking. Investors are mapped in a two-dimensional space $(s^*_i, \bar{n}_{t_i^*})$, where 
\begin{equation*}
    \bar{n}_{t_i^*} = n_{t_i^*}\mathbb{I}[d_i > d_{\theta}]
\end{equation*}
and then the two features are normalized to take values in $[0,1]$. The Euclidean distance between each investor and the point $(1,0)$ is the metric for our ranking. The smaller the distance the higher the ranking.

Finally, it is important to point out the extreme unsupervised nature of our problem. We are not provided with labels associated to each investor so, we train models for dimensionality reduction by using all data and we cannot compare models' performances in terms of accuracy. This and our previous work \cite{our_paper}  tackle the same issue and we could be tempted to employ \cite{our_paper}'s results as ground truth. However, with this dimensionality approach, we aim to provide a new method that could give another tool to support regulators' investigations related to insider trading detection. Therefore, the results of \cite{our_paper} are not validation data: in this new work, they are employed for comparisons and robustness checks.

\subsection{Dimensionality reduction methods}\label{dim_reduction_methods}
\subsubsection{Principal Component Analysis}
\noindent A standard method to apply dimensionality reduction is Principal Component Analysis (PCA) \cite{pca_book}. Starting from a feature scaled data set $X$, the goal of PCA is to obtain a compressed representation $Z_K$ of data and then, a reconstructed version $\hat{X}$ by means of orthogonal transformations: 
\begin{equation*}
    X \rightarrow Z_K = XP_K \rightarrow \hat{X} = Z_KP_K^T 
\end{equation*}
where $X, \hat{X} \in \mathbb{R}^{N,T}, \ Z_K\in \mathbb{R}^{N,K}, \ P_K\in \mathbb{R}^{T,K}$, and the transformation matrix $P_K$ is such that the reconstruction error is minimized with a rank constraint, i.e.
\begin{equation*}
    \hat{X} = \arg\min_{\hat{X}': \  \text{rank}(\hat{X}') \leq K} ||X - \hat{X}' ||_F^2. 
\end{equation*}
The solution is the truncated Singular Value Decomposition (SVD), as follows from the Eckart-Young theorem \cite{eckart1936} or analogously, it is obtained by applying the spectral theorem on the covariance matrix of $X$ which is mean-centered:
\begin{equation*}
    \text{Cov}(X) = \frac{1}{N} X^T X = P \Lambda P^T
\end{equation*}
where $P \in \mathbb{R}^{T,T}$ is orthogonal ($P^T = P^{-1}$), $\Lambda\in \mathbb{R}^{T,T}$ and $\Lambda = \text{diag}(\lambda_1, \lambda_2, \ldots, \lambda_T) $ with $\lambda_1 > \lambda_2> \ldots> \lambda_T$. The sum of the eigenvalues $\lambda_1, \ldots, \lambda_T$ of the covariance matrix is the total variance of the data. Thus, keeping more components means being able to explain more data variability. The eigenvectors matrix $P$ is defined as $P = [p_1, p_2, \ldots, p_T]$ and $p_i$, $i = 1,\ldots, T$ are the loading vectors or principal components. The dimensionality reduction with $K$ components is obtained as 
\begin{equation*}
Z_K = X P_K
\end{equation*}
where $P_K = [p_1, \ldots, p_K] \in \mathbb{R}^{T,K}$, the reconstructed data as $\hat{X} = Z_K P_K^T$ and so $\hat{X} = X P_K P_K^T$. More explicitly, we have
\begin{equation}\label{x_hat_pca}
   \hat{x}_{i}(t) = \sum_{k =1 }^K (x_i \cdot p_k) p_k(t) \ i=1,\ldots, N, \ t= 1, \ldots, T.
\end{equation}.

It is evident that PCA is a decorrelation transformation and that its solution is not unique. Indeed, the loss is invariant under the transformation $P \rightarrow PU$, where $U$ is any orthogonal matrix. Under this transformation, the loading vectors are transformed into a different orthonormal basis for the same subspace. Moreover, according to the Eckart-Young theorem \cite{eckart1936}, the truncated SVD is the best low-rank approximation of a matrix. Therefore, PCA is the linear dimensionality reduction method that minimizes the least squares error in the distortion when we project back to the original space\footnote{Observe that our starting data set is in the format $N\times T$, i.e. the trading days are the features that are subjected to compression. Alternatively, we could start with a data set $Y \in \mathbb{R}^{T,N}$ where the features are the investors. In this case, PCA  consists in identifying a subset of investors, or combinations of them, with a major role in the characterization of agents’ trading activity. However, this would lead to a more time consuming and computationally expensive procedure. Indeed, we would need to obtain the eigendecomposition of the covariance matrix $\text{Cov}(Y) \in \mathbb{R}^{N,N}$, which is much larger than $\text{Cov}(X) \in \mathbb{R}^{T,T}$ since $N \gg T$. Furthermore, as shown in Appendix \ref{appendix_PCA_data_TxN}, if the same feature scaling is applied on the data, the results we obtain starting from the data set in two different formats are analogous.}.

Finally, it is worth noticing that, compared to other insider detection methods, the one based on PCA does not depend on the knowledge of the initial position of the investors.
As explained in Subsection \ref{subsection_method_overview}, this information is indeed lacking in our dataset and in general it can be difficult to obtain because it requires the knowledge of the whole past history of an investor trading activity. Clustering based on k-means, adopted in \cite{our_paper} depends instead on the arbitrary choice of the initial position of the investors.
To show that this is not the case for PCA, we prove that the PCA reconstructed position error $\epsilon$ is invariant under the addition of an arbitrary constant to the vector of positions of a given investor.
{\it Proof.} Let us consider the vector $x_i$ describing the position of investor $i$ and let us add an arbitrary constant $C$ to it, obtaining $x_{i}^C = x_i + C\mathbb{I}_{Tx1}$. Denoting with $\hat{x_i}$ and $\hat{x}_{i}^C$  the reconstructed positions, it is direct to show that $\hat{x_{i}}^C = \hat{x}_i + C\mathbb{I}_{Tx1}$. First, we observe that, given the high number of investors, performing PCA on a dataset where investor $i$ has position $x_i$ and then, on a dataset where investor $i$ has position $x_i^C$ leads to loading vectors which are basically the same. 

Referring to Equation \ref{x_hat_pca},
\begin{equation*}\begin{split}
    \hat{x}_i^C(t)  =& \sum_{k = 1}^K (x_i^C \cdot p_k)p_k(t) = \\
    =&\sum_{k = 1}^K (x_i \cdot p_k)p_k(t) + C\sum_{k = 1}^K (\mathbb{I}_{Tx1} \cdot p_k)p_k(t) = \\
    &= \hat{x_i}(t) + C\sum_{t'=1}^T (P_KP_K^T)_{t't} \\
    &= \hat{x_i}(t) + C \ \ \ \ \ \  \forall t= 1, \ldots, T
\end{split}
\end{equation*}
where $(P_KP_K^T)_{t't}$ is the element $t',t$ of the matrix $P_KP_K^T$ and in the last step we exploit that the matrix $P_K$ is orthogonal. As explained in Subsection \ref{subsection_method_overview}, our anomaly detection approach is \textit{reconstruction-based} and $\epsilon_i = \epsilon_{i}^C$ indeed, 
\begin{equation*}
    \epsilon_{i}^C = || \hat{x_{i}}^C - x_{i}^C || = || \hat{x}_i + C\mathbb{I}_{Tx1} - x_{i} - C\mathbb{I}_{Tx1}|| = \epsilon_{i}.
\end{equation*}
Therefore, the reconstruction error is independent of $C$ and if the profile $x_i$ is identified as anomalous, the same will be true for $x_i^C$. $\square$

However, in our approach, after trading positions are computed, they are normalized according to Equation \ref{eq_norm} and so, computing the position of an investor setting the zero of her portfolio on a different day means that the arbitrary constant $C$ is added to her unnormalized position. Let us define as $\psi_i$ the position of investor $i$ before normalization. Then, we define $\psi_i^C = \psi + C\mathbb{I}_{Tx1}$ as the position of the same investor computed by setting the zero of the portfolio on another day. After normalization, the positions are
\begin{equation*}\begin{split}
    x_i(t)& = \frac{\psi_i(t)}{||\psi_i||_\infty} \\
    x_i^C(t) &= \frac{\psi_i^C(t)}{||\psi_i^C||_\infty} = \frac{\psi_i(t) + C}{||\psi_i + C\mathbb{I}_{Tx1}||_\infty} .
\end{split}
\end{equation*}
Therefore, we have that the reconstructed positions of $x_i^C$ are
\begin{equation*}\begin{split}
    \hat{x}_i^C(t)  = \frac{||\psi_i||_\infty}{||\psi_i + C\mathbb{I}_{Tx1}||_\infty} \hat{x}_i(t) + \frac{C}{||\psi_i + C\mathbb{I}_{Tx1}||_\infty}
\end{split}
\end{equation*}
$\forall t= 1, \ldots, T,$ and the reconstruction error is
\begin{equation*}
    \epsilon_i^C = \frac{||\psi_i||_\infty}{||\psi_i + C\mathbb{I}_{Tx1}||_\infty}  \epsilon_i.
\end{equation*}
This implies that $\epsilon_i^C = \epsilon_i$ if 
\begin{equation}\label{condition_invariance_translation}
    \frac{||\psi_i||_\infty}{||\psi_i + C\mathbb{I}_{Tx1}||_\infty} = 1, 
\end{equation}
which holds if $\max_t{|\psi_i(t)|} = \max_t{ |\psi_i^C(t)|}$ given that $\max_t|\psi_i^C(t)|\neq 0$. As we will show in Section \ref{section_results}, this last condition holds for the majority of the profiles in our dataset.

\subsubsection{Autoencoders}
\noindent PCA is a linear method, consisting in applying the loading vectors' matrix to the starting data twice. Its nonlinear counterpart is an autoencoder (AE)\cite{goodfellow}. Autoencoders' goal is analogous to PCA's: starting from a data set $X$, they aim to obtain a compressed representation of data $Z$ and then, a reconstructed version $\hat{X}$: 
\begin{equation*}
    X \rightarrow  Z = f_1(X)\rightarrow \hat{X}= f_2(Z)
\end{equation*}
where $X, \hat{X} \in \mathbb{R}^{N,T}, \ Z\in \mathbb{R}^{N,K}$. The transformations $f_1$ and $f_2$ are such that the reconstruction error is minimized i.e.
\begin{equation*}
    \hat{X} = \arg\min_{\hat{X}'} ||X - \hat{X}' ||_F^2
\end{equation*}
and, in this case, the compression and decompression steps are made by neural network layers. For an AE with one hidden layer, we have 
\begin{equation}\label{x_hat_ae}
   \hat{x}_{i}(t) = g_2\Big(\sum_{k =1 }^K g_1(x_i \cdot W_1(:,k)) W_2(k,t)\Big)  
\end{equation}
for $i=1,\ldots, N, \ t= 1, \ldots, T$, where $W_1\in \mathbb{R}^{T,K}$, $W_2\in \mathbb{R}^{K,T}$ are layers' weight matrices, also called \textit{encoder} and \textit{decoder}, and $g_1$, $g_2$ are activation functions. The layers' weight matrices are determined by common gradient descent algorithms like Adam \cite{adam}.

Above, we focus on an autoencoder with one hidden layer in order to highlight the analogy with PCA. Indeed, as it is shown in \cite{kunin2019}, if the activation functions are linear, the autoencoder with $L_2$-regularization learn PCA's principal directions. This issue is examined in depth in Appendix \ref{appendix_LAE}. However, autoencoders can be deeper and can have complex architectures such as the well-know convolutional, LSTM, variational autoencoders \cite{goodfellow}.

As we saw above, the solution of PCA follows from the Eckart-Young theorem \cite{eckart1936}. The latter states that the solution to the problem
\begin{equation*}
    \arg\min_{\hat{X}': \ \text{rank}(\hat{X}') \leq K} ||X - \hat{X}' ||_F^2
\end{equation*}
is given by the truncated SVD. The latter approximates excellently data with linear relationships. On the other hand, concerning the reconstruction of nonlinear data, autoencoders outperform PCA, as empirical results in different fields show e.g. applications on image reconstruction. Theoretically, the difference between the problems tackled by PCA and AE is the constraint on the rank of $\hat{X}$. For PCA we impose $\text{rank}(\hat{X}) \leq K$, so there are no more than $K$ columns of $\hat{X}$ which are linearly independent. Equivalently, we have no more than $K$ independent features. On the other hand, AEs aim at minimizing the reconstruction error without any constraint on $\text{rank}(\hat{X})$. This means we could end up with $\text{rank}(\hat{X}) \in (K,T]$ so, more independent features. 

\subsubsection{Pros and cons of the different methods}
\noindent PCA should be preferred against AEs if small datasets are considered, more interpretability and nested solutions are needed. PCA is also easier to implement than AEs. Moreover, it needs less computational resources and less training time. 

The interpretability of the results which are provided by PCA, is due to the linearity of the method. However, this linearity can be a downside if data are nonlinear. On the other end, AEs' nonlinearity allows to capture complex relationships in the data. This last bright side of AEs leads to a better performance in the reconstruction of outliers, compared to PCA, and this could become a downside for our anomaly detection task, if anomalies are reconstructed such that they are indistinguishable from normal data instances.

Finally,  PCA loading factors are ordered such that the associated eigenvalues are in decreasing order: $\lambda_1 \ge \lambda_2 \ge \ldots \ge \lambda_T$. Thus one can measure the importance of a factor in explaining the data and rank the factors according to this criterion. On the contrary, the weight vectors learned by AE are not constrained to form an orthonormal basis, nor to
have a meaningful ordering.

\subsubsection{Choice of K}
\noindent The choice of the dimension $K$ of the compressed representation of the input data should achieve a trade-off between  capturing enough information and avoiding overfitting, which could lead to reconstruct profiles of investors with anomalous behavior well. 

The relying assumption of our use of the dimensionality reduction approach is that the essential and common characteristics are captured by the lower dimensional space and that they explain a large fraction of data variance. Then, anomalous behavior cannot be reproduced given the compression and decompression, and anomalous observations have higher reconstruction errors than normal ones. However, the unsupervised nature of our case makes it extremely complicated, because our training data are anomaly-contaminated.

As a starting point we rely on standard methods to set the parameter $K$, like the percentage of explained variance and the Scree plot, that is the plot of the eigenvalues as a function of $K$ \cite{pca_book}. However, given their erratic nature, we perform an analysis, which helps us in the choice. Let us define $A_K$ as the set of investors identified as anomalous when $K$ is used as dimension of the latent space. First, we determine the cardinality of $A_K$ as a function of $K$. Then, we study the stability of this set by computing the Jaccard similarity \cite{tibi} between each $A_K$ and $A_{K-1}$. $K$ is set to the lowest value of the interval in which we have stability in our results.

\section{Data}\label{section_data}
\subsection{Transaction reporting database}\label{transaction_data}
\noindent The analysis is based on transaction reports collected by Consob for the Italian stocks, according to the directive \href{https://eur-lex.europa.eu/legal-content/IT/ALL/?uri=CELEX:32004L0039&qid=1435044997184}{2014/65} by European Union, also called MiFID II\footnote{In a nutshell, the MiFIDII/MiFIR regime has introduced new regulations for European financial markets and, among them, the transaction reporting obligation that requires investment firms or intermediaries executing transactions in financial instruments to communicate ``complete and accurate details of such trans- actions to the competent authority as quickly as possible, and no later than the close of the following working day''.}.
The relevant dataset was built aggregating the daily transactions of all investors operating in any of the Italian stocks, in the period from January 1, 2019 to September 30, 2021. In details, the dataset was built according to the following rules: i) all the information related to the identity of individual investors have been anonymized; ii) with reference to each stock (identified by its ISIN code), each data point keeps a record of:

\begin{enumerate}
    \item anonymous identifier of the investor;
    \item type of investors (household: H, investment firm: IF, legal entity: L); 
    \item trading venue of the operation (Borsa Italiana - MTA, London Stock Exchange - LSE, off-exchange, etc.) for a total of $224$ venues;
    \item day of the operation;
    \item buy and sell volumes (in shares);
    \item buy and sell Euro volumes;
    \item number of buy and sell contracts;
    \item price of both the first and the last contracts (if there are more than one contract, otherwise they coincide);
    \item minimum and max prices of contracts (if there are more than one contract, otherwise they coincide);
    \item average price of buy (sell) contracts.
\end{enumerate}
In the period covered by the data set, 2,253,707 investors were observed, operating in 286 Italian stocks. This is the same data set used in our previous paper about insider trading detection \cite{our_paper} and in another work related to the investigation of the trading behavior of Italian investors during the Covid pandemic \cite{deriu}.

\subsection{Price sensitive events database}\label{data_PSE}
In addition to the transaction reporting database, a data set containing several price sensitive events (PSEs) was built; such events, obviously public, had all been analysed by the competent Authority with the aim of market abuse detection, by means of standard analytics methodologies. PSEs are events or a set of circumstances relating to listed companies which, when made public, had a significant impact on the price of the company's shares.

Our focus is on insider dealing in the Italian Stock Exchange. Investors who know in advance when a PSE will occur, can trade in a rewarding manner before the information spreads, thus closing their position after the PSE and making a profit. For instance, if a investor knows a few days before its public announcement that a takeover bid is going to occur for a given stock, they could exploit such information by buying shares of the stock considered. When the takeover bid occurs, the shares' price goes up aligning with the offer price and thus, the informed investor can sell by making a no-risk profit.

PSEs dataset contains a list of takeover bids for a number of stocks. As known, a takeover bid is a public offer made by a physical person or a legal entity who is willing to buy other shareholders' shares at a price higher than the stock market value.
As we saw, takeover bids can be exploited by an informed investor by buying before the event. It is worth mentioning that takeover bids have prolonged effects on the market, thus an insider can make a profit even without closing the position immediately after the announcement. 

Our data report for each PSE the stock, its date, and the time window for insider trading investigation. This period varies depending on the type of PSE, which leads to different definitions of the time at which an information starts to be considered price sensitive. In Table \ref{resPSE}, the PSEs database is displayed.

\begin{table}[]
\begin{center}
    \caption{Price sensitive events data set.} 
    \label{resPSE}
    \begin{tabular}{c|c|c}
         Stock  & PSE date & Investigation period ($\Delta$) \\
         \hline
         IMA  & July 28, 2020 & June 29, 2020 - July 28, 2020 \\
         UBI  & Feb 17, 2020 & Jan 16, 2020 - Feb 17, 2020 \\
         PANARIAGROUP  & Mar 31, 2021 & Mar 1, 2021 - Mar 31, 2021 \\
         CARRARO  & Mar 28, 2021 & Jan 4, 2021 - Mar 28, 2021 \\
         MOLMED  & Mar 17, 2020  & Dec 2, 2019 - Mar 17, 2020 \\
    \end{tabular}
\end{center}
\end{table}

\section{Results}\label{section_results}
\begin{figure}[!t]
    \centering\includegraphics[width=.45\linewidth]{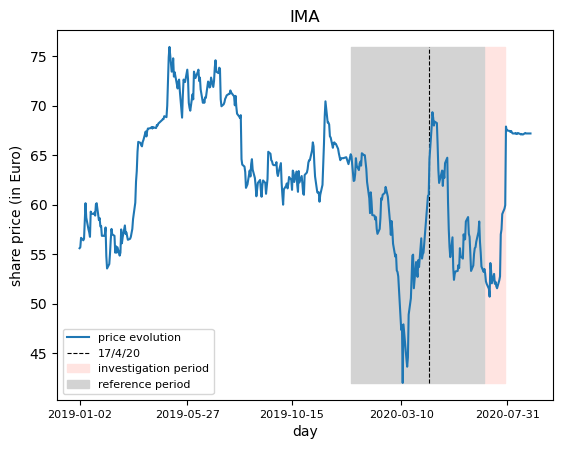}
    \caption{IMA price dynamics. The grey area is the reference period i.e. from January 2, 2020 to June 28, 2020. The pink area is the investigation period i.e. one business month before the PSE.}
    \label{price}
\end{figure}

\noindent As a first case study, we focus on the asset Industria Macchine Automatiche (IMA) whose takeover bid was announced on July 28, 2020.  Figure \ref{price} shows  the price dynamics of this asset. The impact that the PSE had on the share price is evident: there is an increase of $13.16 \%$ on the day of the announcement and the takeover bid's price $68.0$ Euro is reached. In analogy to \cite{our_paper}, we identify the reference period as the time window going from January 2, 2020 to June 28, 2020. Instead, the business month preceding the PSE i.e. July 28, 2020, is the investigation period, as outlined in Table \ref{resPSE}.

For each investor we extract from the database the asset position (in shares) at the end of each day. We assume that the position on January 2, 2020 is zero, but, as proved above, this arbitrary choice has no effect on the reconstruction error.

The trading days are $T = 149$, the investors active and with non-constant position are $N = 13,225$.

\subsection{Anomaly detection with PCA}
\begin{figure}[!t]
    \centering\includegraphics[width=.45\linewidth]{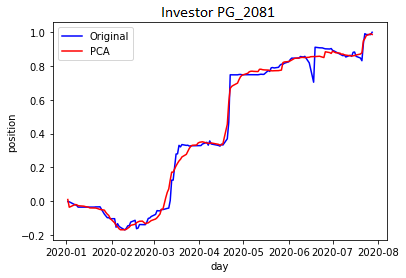}
    \caption{Comparison between the trading position of investor PG\_2081 and the reconstructed one obtained by PCA with $K=16$.}
    \label{trader_PG2081_PCA_vs_original_}
\end{figure}
\noindent The first method we employ in order to perform the dimensionality reduction step, is PCA. In Appendix \ref{app_k} we show the plot of the explained variance as a function of the number of the retained components and the Scree plot. Considering these figures we choose the latent space dimension $K=16$, which allows to retain $97$\% of the explained variance.  
However, we perform a robustness analysis, investigating other choices of $K$ in Appendix \ref{app_k}.

After a feature scaling as pre-processing step, we run our method as illustrated in Section \ref{section_method}. As an example of the reconstructed trading position, in Figure \ref{trader_PG2081_PCA_vs_original_} we show the position of an investor compared to its reconstructed counterpart obtained by PCA. Most of the days the reconstruction is quite close to the original trajectory and the anomaly detection method identifies the days and investors for which the discrepancy, i.e. the reconstruction error, is large.

\begin{figure}[!t]
    \centering\includegraphics[width=.43\linewidth]{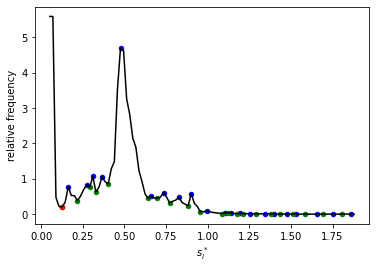}    \includegraphics[width=.46\linewidth]{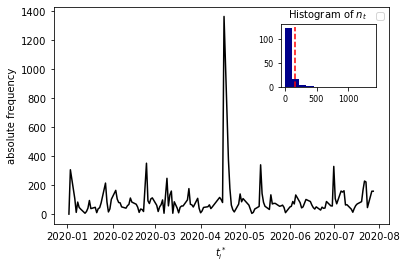}
    \caption{PCA on IMA. Left. Histogram of the anomaly scores. Blue points are local maxima, green local minima and the red point is $\epsilon_\theta \simeq 0.13$ that is the local minimum after the first peak. Right. Histogram of the times when  anomaly scores are observed. The inset plot represents the histogram of the number of investors with given values of the times when their anomaly scores are observed. The vertical dashed line is $n_\theta$ i.e. the top decile of the distribution of $n_t$.}
    \label{anomaly_scores_pdf}
\end{figure}

To identify the thresholds in the anomaly detection method, we plot in Figure \ref{anomaly_scores_pdf}
the histogram of the anomaly score $s_i^*$ (left panel) and of the time of their occurrence $t_i^*$ (right panel). As preannounced, a clear bimodal distribution in the former histogram is observed. The left mode (peak) contains investors with small anomaly score, thus “normal" investors, while the right mode contains potentially anomalous investors with a large maximal reconstruction error. Based on this empirical evidence, we set  $\epsilon_\theta = 0.13$
as the first threshold parameter to identify potential insiders.

By focusing on the histogram of the times $t_i^*$ corresponding to the largest reconstruction errors (right panel of Figure \ref{anomaly_scores_pdf}), we observe the presence of several large peaks, i.e. days when a large number of investors displayed a large reconstruction error. We can understand the origin of these peaks by focusing on the largest one, happened on  April 17, 2020 when more than $1,300$ investors display the largest reconstruction error. Looking at the price dynamics in Figure \ref{price}, we observe that on April 17, 2020 there was a high increase ($7.4\%$) of the share price. 
The large number of investors having the largest reconstruction error on that day is likely due to their reaction to this very volatile day\footnote{We remind that positions of investors are measured in shares and not in Euro, so the peaks are not associated with change in value of a position, due to the large price variation, but to a genuine trading activity.}. Clearly these peaks and the corresponding investors are not insiders and this explains why we impose the condition on $n_t$, that are the heights of the peaks in
the histogram of $t^*_i$, in our methodology to identify anomalous investors - see Equation \ref{eq_anom}. The parameter $n_\theta$ is set equal to $158$, which corresponds to the top decile of the distribution of $n_t$, which cuts off the investors whose anomaly score falls in a peak of $t_i^*$.

Applying our anomaly detection method, we obtain $1,246$ potential insiders out of the $13,225$ total investors. Given the high percentage of anomalous investors that we obtain, their ranking, following the procedure explained at the end of Section \ref{subsection_method_overview}, is fundamental to provide more insight. 

We compare our results with the findings of a method based on k-means similar to that of our previous paper \cite{our_paper}. For each investor we extract the \textit{signed turnover} and the \textit{maximum exposure} in the period\footnote{The \textit{signed turnover} is the aggregated Euro turnover of operations within the period, with positive (negative) sign for a net buying (selling) volume. The \textit{maximum exposure} is the maximum of the absolute value of the position in Euro turnover within the period, with positive (negative) sign if the maximum is reached for a buying (selling) position. We refer to our previous paper \cite{our_paper} for a precise definition.} and we use them as coordinates in a 2D space\footnote{In \cite{our_paper} we consider another feature, namely the \textit{magnitudo/portfolio concentration} which represents the fraction of wealth in the investigated asset. Since our dimensionality reduction method considers only data related to the investigated asset, in the comparison we use a k-means approach in a 2D space.}· Then, we apply k-means to the set of points to identify clusters of investors and we label as anomalous an agent who, in the investigation period belongs to a different cluster than the ones in the reference period and the new cluster is the most rewarding one with respect to the PSE. If the PSE is the announcement of a takeover bid, the cluster with the most rewarding position is the closest to the point $(1,1)$ i.e. both \textit{signed turnover} and \textit{maximum exposure} equal to $1$.
We distinguish two types of anomalous investors. They are \textit{soft} if they are active in the reference period but with a different position than the one in the investigation period, while they are \textit{hard} if they are only active in the investigation period\footnote{See our previous paper \cite{our_paper} for further details.}. 
In summary, the PCA (and later the AE) method acts directly on the whole trading profile of each investor (thus a vector of dimension $T$), while the method based on k-means considers two features, which are functions of the trading profile.

If we run the method based on k-means in 2D on IMA, $152$ investors are identified as \textit{soft} and $705$ as \textit{hard}. Among the $1,246$ potential insiders identified by the method based on PCA, $134$ are \textit{soft} and $671$ are \textit{hard}. From the comparison we find that the first $10$ ranked anomalous investors, according to the method based on PCA, are all identified as suspicious by the method based on k-means. If we consider up to rank $50, 100, 150, 200, 300, 500$, investors who are also suspicious in the framework of the clustering method of \cite{our_paper} are $49, 99, 148, 185, 253, 451$ respectively.

\begin{figure}[!t]
    \includegraphics[width=.45\linewidth]{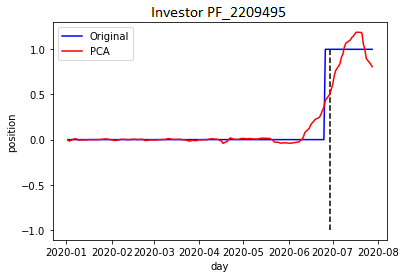}\includegraphics[width=.45\linewidth]{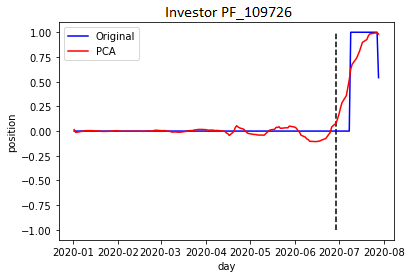}\\   \includegraphics[width=.45\linewidth]{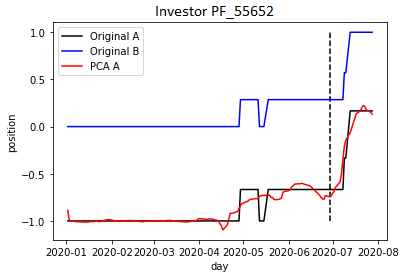}   \includegraphics[width=.45\linewidth]{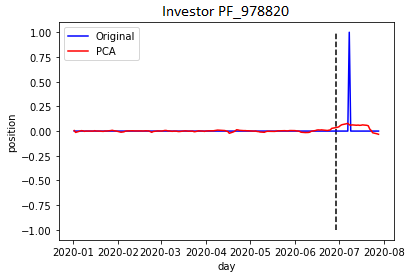}
    \caption{PCA on IMA.  Top panels. Two anomalous investors identified by PCA but not by k-means. Bottom left.  A refers to the trading position of an investor computed setting $t_0$ = January 2, 2019. B refers to the trading position of the same investor computed setting $t_0$ = January 2, 2020. Bottom right. Investor detected by the method based on k-means and not by the method based on PCA. In all panels the vertical dashed line is the day corresponding to the beginning of the investigation period i.e. June 29, 2020.}
    \label{anom_trader_167}
\end{figure}

The compatibility of the two methods is a positive sign of their robustness. However it is natural to ask what are the characteristics of the investors identified as anomalous only by one of the two methods.  
Of the first $500$ ranked anomalous investors, $451$ are also identified by the method based on k-means. Among the remaining ones, $46$ are of the type represented in the top left panel of Figure \ref{anom_trader_167} and $3$ are of the type represented in the top right panel of Figure \ref{anom_trader_167}. The former performs one transaction on June 26, 2020 i.e. three days before the starting day of the investigation period. This investor could be suspicious given her aggressive buying position just in the vicinity of the PSE. However, this investor is not identified by the method based on k-means since the transaction is outside the investigation period. 
Therefore, contrary to the method based on k-means, the new method based on a dimensionality approach is not strictly dependent on an arbitrary choice of the investigation period.

On the other hand, the investor in the top right panel of Figure \ref{anom_trader_167} sells a portion of her position on the day before the PSE, but still maintains a net buying position. Given this investor was not active in the reference period, this behavior of buying and then, selling in the investigation period, could be a strategy in order not to be identified as suspicious by the regulator. This investor is not identified by the method based on k-means since in the last time window there is a drop in her \textit{signed turnover} which leads the corresponding point to move away from the most rewarding cluster. 

Another positive aspect of the new method based on a dimensionality reduction approach is its ability to detect given investors as anomalous, independently of the starting point chosen for the computation of her position (see Subsection \ref{dim_reduction_methods}). We observe that the condition in Equation \ref{condition_invariance_translation} - which states when the reconstruction error of a given profile is the same of its shifted counterpart - holds for about $63\%$ of investors in our dataset, if January 2, 2020 and January 2, 2019 are two choices of zeros for the computation of the positions. In the bottom left panel of Figure \ref{anom_trader_167} 
we show the position of one investor when setting to zero the position on January 2, 2020 (blue) or on January 2, 2019 (black). 
Interestingly, this investor is identified as anomalous by the method based on PCA, but not by the method based on k-means. The fact that with the choice $t_0= \text{January} \ 2,\ 2019$  this investor is identified as anomalous could surprise given we are focusing on detecting insider trading related to the announcement of a takeover bid. We recall that after this kind of PSE, the price increases and so, insiders are likely to have positive positions before the PSE. Indeed, this investor is not identified as anomalous by the method based on k-means given her negative value of \textit{signed turnover}. On the other hand, if the profile is computed with the choice $t_0 = \text{January} \ 2, \ 2020$, she has a net buying position before the PSE and \textit{signed turnover} equal to $1$. Therefore, only with this choice of $t_0$, she is identified as \textit{soft discontinuous} by the method based on k-means.

It is also interesting to investigate why some investors are detected by the method based on k-means and not by the method based on PCA. The total number of these investors is $79$; among them, $27$ are investors with constant position in terms of shares and so, they are not included in the analysis  based on PCA. The majority of the remaining profiles are of the type displayed in the bottom right part of Figure \ref{anom_trader_167}. They are investors who have a null position in the investigation period, if positions are computed in shares. Instead, if positions are computed in Euros, as in the method based on k-means, the \textit{signed turnover} and the \textit{maximum exposure} of these investors are equal to $1$ in the investigation period. They are in the best rewarding position and this makes them extremely suspicious according to the clustering approach of \cite{our_paper}. On the contrary, these investors are not identified by the method based on PCA, since the condition (d) relative to Equation \ref{eq_anom} does not hold. That condition requires that the difference between the position
on the PSE and the position on the first day of the reference period is larger than 0.5. For investors like PF\_978820 (bottom right part of Figure \ref{anom_trader_167}), this difference  is null since the investor closes her position before the PSE.

Finally, in Appendix \ref{appendix_PCA_data_TxN}, we provide a comparison between the results obtained starting with the data set in the formats $N\times T$ and $T\times N$. As shown in Figure \ref{eigenvaluesPCA_NxT_vs_TxN.png} and \ref{anomaly_score_pdf_NxT_vs_TxN}, if data are feature scaled in the same way, there is no difference between these results. In Appendix \ref{appendix_LAE}, the relation between PCA and $L_2$-regularized autoencoders is tested on our data set.

\subsection{Going nonlinear: the Autoencoder}
\noindent As extensively proved in other research fields such as image reconstruction, adopting nonlinear and deep autoencoders can lead to a gain in performances, giving their ability to capture more complex relations in data. However, it is important to stress that our ultimate goal is not to best reconstruct our data. We wish to achieve a trade-off and to avoid overfitting. The idea is to obtain a lower dimensional space which captures the common and essential data characteristics; in this way, normal trading profiles will be well described while anomalous ones will not. 

\begin{table}
    \begin{center}
    \caption{Autoencoders' architectures.}\label{tab_architectures}
        \begin{tabular}{c|c|c|c}
         Name &  Neurons of the  &  Encoding & Neurons of the \\
         & hidden layers  &dimension & hidden layers  \\
         & in the encoder &  & in the decoder \\
         \hline
         AE-1 & - & 16 & - \\
         AE-2 & 32 & 16 & 32 \\
         AE-3  & 64, 32 & 16 & 32, 64\\
         AE-4  & 128, 64, 32 & 16 &  32, 64, 128 
    \end{tabular}
    \end{center}
\end{table}

\begin{figure}[!t]
    \centering\includegraphics[width=.45\linewidth]{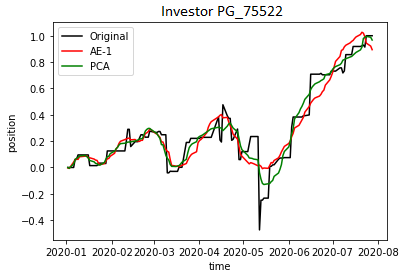}
    \caption{IMA. Comparison between the trading position of investor PG\_75522 and the reconstructed ones obtained by PCA and AE-1 with $K=16$.}
    \label{trader_PG75522}
\end{figure}

Bearing this in mind, we investigate the use of nonlinear autoencoders for our problem. The symmetric architectures we employ are schematized in Table \ref{tab_architectures}. 
The number of neurons is chosen according to the geometric pyramid rule \cite{pyramid_rule} and for all architectures the activation function of the hidden layers is the ReLU, while the activation function of the output layer is the hyperbolic tangent. This last choice allows to exploit the nonlinearity of the neural networks and yet to  produce outputs with values in the interval $[-1,1]$, the domain of the normalized trading positions. The loss function is the mean squared error (MSE) and Adam \cite{adam} is used as optimization algorithm. 

\begin{table*}
    \begin{center}
    \caption{Metrics for different dimensionality reductions of the IMA dataset. The results related to the autoencoders are averaged over $10$ runs.}
    \label{tab_comparison_models_IMA}
       \begin{tabular}{c|c|c|c|c|c|c|c}
    Model & $||X - \hat{X}||_F$ & EVS & $\bar{s^*}$ & $\bar{s^*}_{anomalous}$ & $\bar{s^*}_{normal}$ & $\text{M}_1$ & $\text{M}_2$ \\
    \hline
     PCA & 132.0 & 97.66 &0.4525 & 0.5051 & 0.4489 & 0.1251 & 1.116 \\
     AE-1 & 168.7 & 96.18 &0.5100 & 0.6088 & 0.5031 & 0.2099 &  1.193\\
     AE-2 & 137.5 & 97.46 & 0.4574 & 0.5128 & 0.4536 & 0.1308 & 1.121  \\
     AE-3  & 129.3 & 97.76& 0.4404 & 0.4997 & 0.4363 & 0.1453 & 1.134\\
     AE-4 & 121.1& 98.00 & 0.4228 & 0.4620 & 0.4201 & 0.0999 & 1.093
    \end{tabular} 
    \end{center}
\end{table*}

The PCA and the $4$ autoencoder architectures of Table \ref{tab_architectures} are run on our data set of trading positions for IMA. A comparison between the reconstructed profile of an investor, obtained with PCA and a type of autoencoder is represented in Figure \ref{trader_PG75522}. Greater smoothness is associated to the profile obtained by AE-1 however, the overall similarity between the two different profile reconstructions is evident. Table \ref{tab_comparison_models_IMA} summarizes the main results for all autoencoders' architectures in terms of several metrics. Due to their nonlinear character, we expect deep autoencoders can capture more complex features in data. This could lead to the identification of characteristics for the data compression which are more relevant than the ones identified by PCA, which is a linear model.
However, we need at least $5$ hidden layers to outperform PCA in terms of the loss function $||X-\hat{X}||_F$. The lower MSE is accompanied by greater explained variance score (EVS) which is defined as\footnote{Notice that the EVS is different from a common metric that is usually employed for PCA, that is the explained variance ratio (EVR), defined as
\begin{equation*}
    \text{EVR} = \frac{\sum_{k=1}^K \text{var}(z_k)}{\sum_{t=1}^T \text{var}(x_t)},
\end{equation*}
where $x_t \in \mathbb{R}^{N}$ are the columns of $X\in \mathbb{R}^{N,T}$, that is the data matrix, and $z_k \in \mathbb{R}^{N}$ are the columns of $Z_K\in \mathbb{R}^{N,K}$, that is $X$'s representation in the $K$-dimensional latent space. The EVR is not a good indicator for autoencoders since it strictly depends on the activation functions we choose. Therefore, EVS is preferred to EVR, thus allowing a fair comparison with the PCA results.}
\begin{equation*}
    \text{EVS} = 100\Bigg(1 - \frac{\text{Var}(X - \hat{X})}{\text{Var}(X)}\Bigg).
\end{equation*}
This means that deeper autoencoders can explain a larger variance of the data.

Nonetheless, we are not solely interested in better reconstruction errors. A larger gap between the errors of the anomalous investors and the ones who are not anomalous is desired. Given our unavailability of labels, we compare the mean anomaly score of investors who are detected as \textit{hard/soft} by the method based on k-means i.e. $\bar{s^*}_{anomalous}$, and the mean anomaly score of the other investors i.e. $\bar{s^*}_{normal}$.

The two metrics $\text{M}_1$ and $\text{M}_2$, that we introduce, shed light on this issue. They are defined as 
\begin{equation*}
    \text{M}_1 =\frac{\bar{s^*}_{anomalous} - \bar{s^*}_{normal}}{\bar{s^*}_{normal}}, \
    \text{M}_2 = \frac{\bar{s^*}_{anomalous}}{\bar{s^*}}.
\end{equation*}

We obtain that the model AE-1 has the greatest values of $\text{M}_1$ and $\text{M}_2$ i.e. it leads to a greater gap between the anomaly scores of our proxy of anomalous investors and the others. On the other hand, AE-4, which allows to obtain the lowest error in the data reconstruction, leads to the lowest
value of the two quantities. It is important to stress that the comparison between the values of $\text{M}_1$ and $\text{M}_2$ that are obtained with different models, provide information which could be useful to our insiders detection task. However, the anomaly score is not the only quantity which determines whether an investor is identified as anomalous; the distribution of the times corresponding to the largest reconstruction errors also plays a fundamental role, both on the identification and on the ranking. We will investigate this issue in the following by employing both AE-1 and AE-4 for the anomaly detection step.

\subsection{Anomaly detection with autoencoders}

\begin{table*}
    \begin{center}
    \caption{IMA: anomaly detection. $A$ is the set of investors identified as anomalous. $I$ is defined as $I= |A \cap A^{KM}|$ where $A_{KM}$ is set of investors identified as \textit{hard/soft} by the method based on k-means. $I_n$ is defined as $I_n = |A_{n} \cap A^{KM}|$ where $A_{n}$ is the set of the first $n$ ranked anomalous investors.}
    \label{tab_IMA_AE}
      \begin{tabular}{c|c|c|c|c|c|c|c|c|c}
    Method & $|A|$ & $I/|A_{KM}|$ & $I_{10}$ & $I_{50}$ & $I_{100}$ & $I_{150}$ & $I_{200}$ & $I_{300}$ & $I_{500}$ \\
    \hline
        PCA &  1,246 & 805/857 & 10 & 49 & 99 & 148 & 185 & 253 & 451\\
        AE-1 & 1,502 & 812/857 & 10 & 49 & 99 & 148 & 186 & 193 & 337\\
        AE-4 & 1,325 & 807/857 & 10 & 49 & 99 & 148 & 166 & 226 & 424 
    \end{tabular}  
    \end{center}
\end{table*}

\noindent We perform our anomaly detection task by employing two different architectures of autoencoders i.e AE-1 and AE-4. 
The main results are summarized in Table \ref{tab_IMA_AE} and compared with the results of PCA.

First of all we notice that when considering the first $150$ ranked investors, the different methods provides almost identical set of anomalous cases. This, once more, indicates that machine learning methods (k-means, PCA, autoencoders) essentially agree in the identification of the most suspicious investors, demonstrating an overall robustness of the adopted methodologies.

\begin{figure}[!t]
    \includegraphics[width=.415\linewidth]{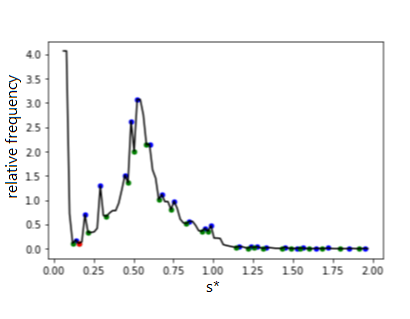}   \includegraphics[width=.45\linewidth]{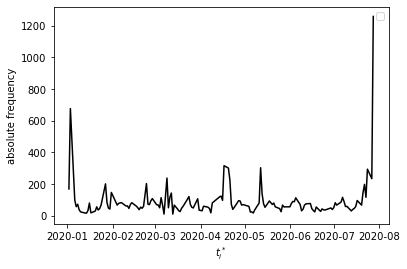}
    \caption{AE-1 on IMA.  Left. Histogram of the anomaly scores; blue points are local maxima, green local minima and the red point is $\epsilon_\theta \simeq 0.156$ that is the local minimum after the first peak. Right. Histogram of the times corresponding to the anomaly scores.} \label{IMA_anomaly_score_pdf_AE_1.png}
\end{figure}

When  we focus on the first $500$ ranked investors, PCA is the method with the largest overlap with the method based on k-means. This cannot be explained by the anomaly score values since, as shown in Table \ref{tab_comparison_models_IMA}, the value of the metrics $\text{M}_1$, $\text{M}_2$ obtained by PCA are lower than the one obtained by AE-1. The two different dimensionality reduction methods lead to different histograms of the times corresponding to the largest reconstruction errors, which are shown in the right panels of Figure \ref{anomaly_scores_pdf} and Figure \ref{IMA_anomaly_score_pdf_AE_1.png} (Appendix \ref{ad_ae}). While the histogram obtained with PCA has a maximum on April 17, 2020, the one obtained with AE-1 has the highest peak on the day of the PSE. Contrary to PCA, the autoencoder AE-1 is able to provide a dimensionality reduction where the trading activity on April 17, 2020 is treated as “normal" for a large fraction of investors. Therefore, the decrease of the overlapping could be ascribed to the change in the distribution of the times corresponding to the largest anomaly scores.

Among the first $500$ ranked investors identified by AE-1 or AE-4, investors who are detected by AE-1 or AE-4 and that are not detected by the method based on k-means are analogous to the ones which were identified by relying on PCA and not by the method based on k-means i.e. profiles like the one in the top left panel of Figure \ref{anom_trader_167}, with one buying transaction just before the beginning of the investigation period. 

\begin{figure}[!t]
    \centering\includegraphics[width=.45\linewidth]{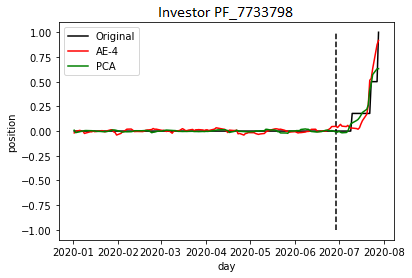}
    \caption{IMA. Investor detected by the method based on AE-4 and not by the
method based on PCA. The vertical dashed line is the day corresponding to the beginning of the investigation period i.e. June 29, 2020.}
\label{PF_7733798_AE_4}
\end{figure}

Instead, among the first $500$ ranked investors identified by AE-1, $62$ are not identified by PCA. Among them, the investors ranked $15$ (Figure \ref{PG_1903444_AE_1} in Appendix \ref{ad_ae}), $114, 146, 396$ are identified as \textit{hard} by the method based on k-means. The others are of the type represented in the top left panel of Figure \ref{anom_trader_167}. In a similar way, among the first $500$ ranked identified by AE-4, investors who are not detected by PCA are $16$. Among them, $3$ are \textit{hard} and are like the investor in Figure \ref{PF_7733798_AE_4} (Appendix \ref{ad_ae}), the others are like the one in the top left panel of Figure \ref{anom_trader_167}. It is evident the ability of the autoencoders to capture as anomalous a type of profile like the ones in Figure \ref{PG_1903444_AE_1} and \ref{PF_7733798_AE_4} (Appendix \ref{ad_ae}), that were not identified by PCA and that are \textit{hard} according to the method based on k-means. Moreover, again, the method based on dimensionality reduction approaches shows its independence of the choice of the investigation period.

If we compare the results obtained by employing different architectures of autoencoders, among the first $500$ ranked by AE-4, only $1$ was not detected by AE-1. 

\subsection{Other case studies}
\begin{table*}
\begin{center}
\caption{Anomaly detection on all assets, obtained by employing PCA. $N$ and $T$ are the numbers of investors and days in the data set. $K$ is the encoding dimension. $A$ is the set of investors identified as anomalous. $I$ is defined as $I= |A \cap A^{KM}|$ where $A_{KM}$ is set of investors identified as \textit{hard/soft} by the method based on k-means. $I_n$ is defined as $I_n = |A_{n} \cap A^{KM}|$ where $A_{n}$ is the set of the first $n$ ranked anomalous investors.}
    \label{tab_all_PCA}
    \begin{tabular}{c|c|c|c|c|c|c|c|c|c|c|c|c}
    Asset & N & T & $K$ & $|A|$ & $I/|A^{KM}|$ &  $I_{10}$ & $I_{50}$ & $I_{100}$ & $I_{150}$ & $I_{200}$ & $I_{300}$ & $I_{500}$ \\
    \hline
        IMA & 13,225 & 149 & 16 & 1,246 & 805/857 & 10 & 49 & 99 & 148 & 185 & 253 & 451 \\
        UBI  & 31,970 & 118 & 16  & 1,801 & 1,255/1,432 & 10 & 50 & 100 & 150 & 200 & 300 & 499\\
        PANARIAGROUP & 1,068 & 56 & 12 & 232 & 178/188 & 10 & 42 & 91 & 125 & 150 & - & -\\
        CARRARO & 4,500 & 317 & 24 & 537 & 431/500 & 9 & 49 & 99 & 149 & 199 & 283 & 401\\
        MOLMED &  11,976 & 307 & 38 & 1121 & 465/1,264 & 1 & 10 & 41 & 60 & 62 & 86 & 286 \\
    \end{tabular}
\end{center}
\end{table*}

\begin{table*}
\begin{center}
\caption{UBI: dimensionality reduction. The results related to the autoencoders are averaged over $10$ runs.}
    \label{tab_comparison_models_UBI}
    \begin{tabular}{c|c|c|c|c|c|c|c}
    Model & $||X - \hat{X}||_F$ & EVS & $\bar{s^*}$ & $\bar{s^*}_{anomalous}$ & $\bar{s^*}_{normal}$ & $\text{M}_1$ & $\text{M}_2$ \\
    \hline
     PCA &  205.8 & 97.07 & 0.4953 & 0.5982 & 0.4905 & 0.2196 & 1.208\\
     AE-1 & 234.0 & 96.20 & 0.5225 & 0.6123 & 0.5183 & 0.1816 & 1.172 \\
     AE-2 &  185.6 & 97.61 & 0.4417 & 0.5267 & 0.4377 & 0.2034 & 1.192  \\
     AE-3  & 168.3 & 98.04 & 0.4003 & 0.5015 & 0.3955 & 0.2684 & 1.253\\
     \end{tabular}
\end{center}
\end{table*}

\noindent While we have extensively covered the case study related to the asset IMA, we now focus on the other PSEs shown in Table \ref{resPSE}. Table \ref{tab_all_PCA} summarizes the main results obtained by using PCA. The overlapping with the results of the method based on k-means are analogous to what is obtained for IMA, except for MOLMED. For this asset, the small overlapping is due to the choice of the investigation period, thus highlighting the ability of our method based on a dimensionality reduction approach to be independent of the choice of the investigation period. Moreover, the value added by our new method to the insider trading detection task, is analogous to what is obtained for IMA.

Now, let us deepen the main results related to the asset UBI. We employ both PCA and autoencoders. In Table \ref{tab_comparison_models_UBI}, a comparison between the reconstruction results obtained by employing different architectures is shown. A trend different from IMA can be observed. In this case, we need at least $3$ hidden layers to outperform PCA in the reconstruction of trading profiles and AE-3 shows the lowest MSE. Contrary to IMA, the architecture which leads to the greatest values of $\text{M}_1$ and $\text{M}_2$ is still AE-3. If we rely on this autoencoder and apply our anomaly detection method, the results we obtain are provided in Table \ref{tab_AD_UBI}, compared with the ones of PCA.

\begin{table*}
\begin{center}
\caption{UBI: anomaly detection. The encoding dimension is $16$. $A$ is the set of investors identified as anomalous. $I$ is defined as $I= |A \cap A^{KM}|$ where $A_{KM}$ is set of investors identified as \textit{hard/soft} by the method based on k-means. $I_n$ is defined as $I_n = |A_{n} \cap A^{KM}|$ where $A_{n}$ is the set of the first $n$ ranked anomalous investors.}
    \label{tab_AD_UBI}
    \begin{tabular}{c|c|c|c|c|c|c|c|c|c}
    Method & $|A|$ & $I/|A_{KM}|$ & $I_{10}$ & $I_{50}$ & $I_{100}$ & $I_{150}$ & $I_{200}$ & $I_{300}$ & $I_{500}$ \\
    \hline
         PCA &  1,801 & 1,255/1,432 & 10 & 50 & 100 & 150 & 200 & 300 & 499\\
         AE-3& 2,106 & 1,348/1,432 & 10 & 50 & 100 & 150 & 200 & 300 & 457
    \end{tabular}
\end{center}
\end{table*}

If the first $500$ ranked by PCA are considered, our new method does not provide new information compared to the one based on k-means. This is in contrast with the autoencoder. The profiles detected by AE-3 and not by k-means are analogous to the one in the top left panel of Figure \ref{anom_trader_167}. On the other hand, the investors detected by AE-3 and not by PCA are \textit{hard} according to the method based on k-means and analogous to the profile in Figure \ref{PF_7733798_AE_4} (Appendix \ref{ad_ae}). The ability of autoencoders to capture this type of investors which are not detected by PCA, was already shown in the study related to IMA. 

\section{Conclusion}\label{section_conclusion}
\noindent We proposed a novel unsupervised approach for contextual anomaly detection, to support decision in insider trading detection. This method tackles the same issue of our previous paper \cite{our_paper} with a different point of view. In particular, the method based on k-means, that we develop in \cite{our_paper}, is based on the definition of three features i.e. \textit{signed turnover}, \textit{magnitudo}, \textit{maximum exposure}. With this new method, we aim at overcoming the features' choice: our only input is the trading position of each investor for a given asset and the model learns the relevant characteristics by itself. 

This new approach lies in the \textit{reconstruction-based} paradigm of anomaly detection and it involves several steps. First, we employ PCA or autoencoders and we obtain the reconstruction errors for the trading profiles of each investor active on a given asset for which we have a takeover bid. Then, we localize the largest errors and impose several conditions in order to detect anomalous investors, who could be suspicious of insider trading related to the PSE.

We observe a consistent overlapping with the results of the method based on k-means. However, the value added of this new method is evident. If PCA is employed as dimensionality reduction approach, the method is extremely fast and easy to implement. Both with PCA and autoencoders, we do not longer have to choose the trading features which allow to characterize the trading activity of each investor. The method is not strictly dependent on the choice of the beginning of the investigation period and actually, it could provide insight on whether this time window should be fixed. The method is also independent of the choice of the initial time for the computation of the trading positions. 

The differences between the performances of PCA and autoencoders are case-by-case dependent. We showed that autoencoders allow to identify as anomalous, profiles that are not detected by PCA and are actually \textit{hard} according to the method based on k-means. We think that for small data sets, PCA is a sufficient method to perform the dimensionality reduction step. Instead, for larger data sets, a coupled use of PCA and autoencoders should be preferred. This conclusion is also motivated by the extreme complexity of our problem, that is also strengthened by the unavailability of labels, which force us to evaluate the performance of our method without a systematic procedure.

A natural extension of this work is the employment of more complex architectures of autoencoders.

\section*{Declarations}
\noindent This paper represents the personal opinions of the authors and does not bind the membership organization in any way.

Restrictions apply to the availability of the data, because of the severe privacy policy related to the data collected within the MiFIDII/MiFIR regime and are not publicly available.

\appendices
\section{The choice of $K$ in PCA}\label{app_k}
\noindent In this Appendix we investigate the optimal dimension $K$ in PCA analysis and we perform a robustness check. The results refer to IMA.

\begin{figure}[!t]
    \includegraphics[width=.45\linewidth]{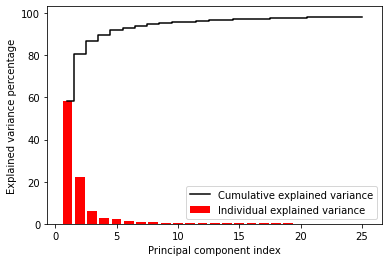}  \includegraphics[width=.45\linewidth]{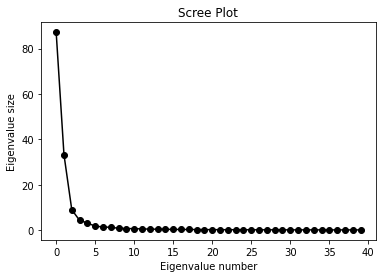}
    \caption{PCA on IMA. Explained variance percentage as a function of the number of retained components (left) and Scree plot i.e. eigenvalue size as a function of the corresponding component index (right).}
    \label{pca_exp_var}
\end{figure}

The explained variance percentage and Scree plots are shown in Figure \ref{pca_exp_var}. Keeping only $1$ component allows to retain $58.5 \%$ of the data variance and we only need $5$ components to reach $90 \%$.

\begin{figure}[!t]
    \centering\includegraphics[width=.45\linewidth]{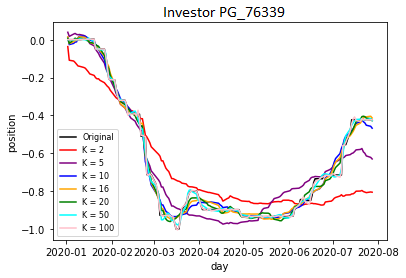}
    \caption{Comparison between the trading position of investor PG\_76339 and the reconstructed ones obtained by PCA with several values of $K$.}
    \label{trader_PG76339_PCA_vs_original_severalK}
\end{figure}

In Figure \ref{trader_PG76339_PCA_vs_original_severalK}, the trading profile of an investor is compared with the reconstructed ones obtained by PCA for several values of $K$. It is evident that increasing the latent space dimension leads to an improvement in the reconstruction of the profile. Moreover, $K=16$ allows to obtain reconstruction errors which are comparable to the ones related to higher values of $K$. Indeed, the choice $K=16$ is such that $97\%$ of the data variance can be explained. However, as we illustrated in Subsection \ref{dim_reduction_methods}, giving the unsupervised nature of our problem and our complete lack of labels, \textit{a priori} we do not know the best suited value of this parameter. This motivates us to perform an analysis to understand how different choices of $K$ could impact our results.

\begin{figure}[!t]
    \includegraphics[width=.45\linewidth]{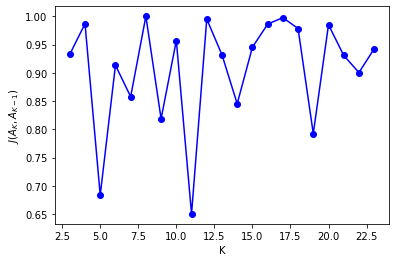}    \includegraphics[width=.45\linewidth]{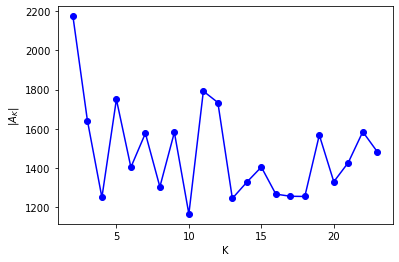}
    \caption{Left. Jaccard similarity between the set of anomalous investors identified with $K$ and $K-1$. Right. Cardinality of the set of anomalous investors as a function of $K$.}
    \label{jaccard}
\end{figure}

For several choices of $K$, we run our methodology using PCA for the dimensionality reduction step. Then, we identify a set of anomalous investors $A_K$ and in order to test the stability of this set, we compute the Jaccard similarity \cite{tibi} between each $A_K$ and $A_{K-1}$. Results are reported in the left panel of Figure \ref{jaccard}, which shows that, especially for small values of $K$, the similarity is unstable and oscillates. On the other hand, in an interval between $K = 16$ and $K = 18$, we have a more stable trend with very high values of the metric. This stability is also evident by looking at the right panel of Figure \ref{jaccard}, which shows the cardinality of $A_K$ for several values of $K$. This motivates us to set $K=16$ i.e. the lowest value of the interval in which we have stability in our findings.

\section{PCA on data in two different formats}\label{appendix_PCA_data_TxN}
\noindent Our starting data set is in the format $N\times T$ i.e. the trading days are the features which are subjected to compression. We could start with a data set $Y \in \mathbb{R}^{T,N}$ where the features are the investors. As we explained in Subsection \ref{dim_reduction_methods}, this would lead to a more time consuming and computationally expensive procedure since in our dataset $N \gg T$. However, it is legitimate to ask whether there is a difference in the results obtained with these two approaches.

Before running PCA, it is fundamental to perform feature scaling. This preprocessing step consists in rescaling each feature such that it has unit standard deviation and null mean. Our input data are investors' positions which are normalized as explained in the main text. This first normalization is such that the activity of each investor is normalized compared to her own trading history. If the features are the trading days, the feature scaling before PCA leads to a data set where 
\begin{equation}\label{feature_scaling}
    x_{i}(t) \rightarrow \frac{x_{i}(t) - \text{mean}(x_{t})}{\text{std}(x_{t})}
\end{equation}
where $x_t$ are the columns of $X \in \mathbb{R}^{N,T}.$
Therefore, this second normalization step consists in normalizing the position of each investor on a day with respect to the positions of all other investors on that day.

If instead, the feature scaling is performed with respect to investors, it would lead to a data set where 
\begin{equation*}
    x_{i}(t) \rightarrow \frac{x_{i}(t) - \text{mean}(x_{i})}{\text{std}(x_{i})}.
\end{equation*}
where $x_i$ are the rows of $X \in \mathbb{R}^{N,T}.$
This means we are normalizing the position of each investor on a day with respect to the positions of the same investor on other days. We remind that also the normalization used in the main text, although different, uses the whole history of an investor's position.

\begin{figure}[!t]
   \centering\includegraphics[width=.45\linewidth]{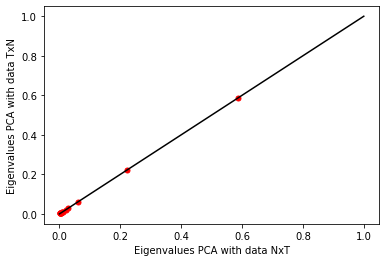}
   \caption{Comparison between the eigenvalues obtained by running PCA on data in the formats $N\times T$ and $T\times N$. The dark line is the bisector.}
    \label{eigenvaluesPCA_NxT_vs_TxN.png}
\end{figure}

\begin{figure}[!t]
   \centering\includegraphics[width=.42\linewidth]{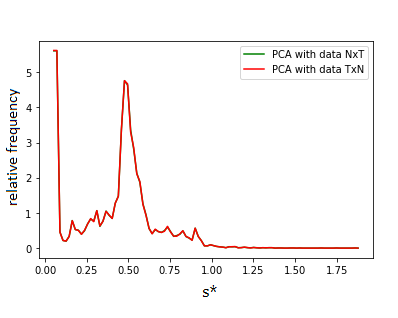}
   \caption{Comparison between the anomaly scores histograms obtained by running PCA on data in the formats $N\times T$ and $T\times N$.}
    \label{anomaly_score_pdf_NxT_vs_TxN}
\end{figure}

Therefore, we adopt the feature scaling of Equation \ref{feature_scaling} and we apply PCA using as input the data set in the format $N\times T$ and then, in the format $T\times N$. The eigenvalues that we obtain in the two cases are the same, as it is represented in Figure \ref{eigenvaluesPCA_NxT_vs_TxN.png}. Analogously, Figure \ref{anomaly_score_pdf_NxT_vs_TxN} shows the equivalency between the anomaly scores distributions.

Formally, this can be explained by observing that PCA identifies the eigenvalues of the data covariance matrix. This means:
\begin{equation*}
    (X^TX)p = \lambda p 
\end{equation*}
where $\lambda$ is an eigenvalue and $p$ is the corresponding eigenvector. If we multiply by $X$, we obtain
\begin{equation*}
    (XX^T)(Xp) = \lambda (X p) \iff \text{Cov}(Y)(X p) = \lambda (X p)
\end{equation*}
where $Y = X^T$. Therefore, the eigenvalues of $\text{Cov}(X)$ and $\text{Cov}(Y)$ are the same while the eigenvectors are $p$ and $X p$.

\section{Relation between linear autoencoders and PCA}\label{appendix_LAE}
\noindent If we compare Equations \ref{x_hat_pca} and \ref{x_hat_ae}, it is pretty evident they are analogous if the activation functions $g_2$ and $g_1$ are the identity functions and $W_1 = W_2^T$. Indeed, as illustrated in \cite{kunin2019}, if the activation functions are linear, the autoencoder with $L_2$-regularization learns PCA's principal directions.

Given the data $X\in \mathbb{R}^{N,T}$, linear autoencoders' (LAEs) goal is to obtain the following transformations: 
\begin{equation*}
    X \rightarrow Z = X W_1 \rightarrow \hat{X} = X W_1 W_2  
\end{equation*}
where $Z\in \mathbb{R}^{N,K}$, $W_1\in \mathbb{R}^{T,K}$, $W_2\in \mathbb{R}^{K,T}$, and such that the loss function is minimized i.e.
\begin{equation*}
    W_{1,2} = \arg\min_{W_1, W_2} \mathcal{L}(W_1, W_2) = \arg\min_{W_1, W_2} ||X - X W_1W_2 ||_F^2.
\end{equation*}
As for standard autoencoders, $W_1$ is called \textit{encoder} and $W_2$ \textit{decoder}.

By the Eckart-Young theorem \cite{eckart1936}, the optimal rank-K solution is the truncated Singular Value Decomposition (SVD) i.e.
\begin{equation*}\begin{split}
    X W_1W_2 &= U_K S_K V_K^T = U S I_{T xK} V_K^T = \\
    &= U S V^T V_K V_K^T = X V_K V_K^T.
\end{split}
\end{equation*}
Therefore, a LAE learns the principal subspace. However, it does not learn the principal directions indeed $W_1, W_2$ are optimal under the following transformations:
\begin{equation*}\begin{split}
    W_1 \rightarrow W_1 G \\
    W_2 \rightarrow G^{-1} W_2   \\
    \forall G \in GL_K(\mathbb{R})\\
\end{split}
\end{equation*}
where $GL_K(\mathbb{R})$ is the general linear group i.e. matrices which are invertible.

Contrary to traditional PCA loading factors, the weight vectors learned by a LAE are not constrained to form an orthonormal basis, nor to
have a meaningful ordering. However, they span the same subspace.

If instead of $\mathcal{L}(W_1,W_2)$, we consider 
\begin{equation*}
    \mathcal{L}_\sigma(W_1,W_2) = \mathcal{L}(W_1,W_2) + \lambda(||W_1||_F^2 + ||W_2||_F^2), \ \lambda > 0, 
\end{equation*}
the penalization term $\lambda(||W_1||_F^2 + ||W_2||_F^2)$ is not invariant under the general linear group indeed
\begin{equation*}
    ||\alpha W_1||_F^2 = \alpha^2 ||W_1||_F^2 \neq ||W_1||_F^2.
\end{equation*}
On the other hand, it is invariant under the orthogonal group indeed
\begin{equation*}
    ||W_1 O||_F^2 = ||W_1||_F^2 \ \forall O \in O_K(\mathbb{R})
\end{equation*}
and we recall $O_K(\mathbb{R}) \subset GL_K(\mathbb{R})$. So, $\mathcal{L}(W_1,W_2)$ is invariant under the general linear group while $\mathcal{L}_\sigma(W_1,W_2)$ under the orthogonal group (the invariance is considered with respect to the transformation applied to $W_1$ and $W_2$).

As we said above, if $W_1$ is optimal, so does $W_1G$ $\forall G \in GL_K(\mathbb{R})$ and we observe that
\begin{equation*}
    W_1 G = US V^T G 
\end{equation*}
i.e. it is not in SVD form. On the other hand, we have that
\begin{equation*}
    W_1 O = US V^T O 
\end{equation*}
i.e. $W_1 O$ is in SVD form.

In \cite{kunin2019}, after this reasoning, authors provide an algorithm to recover the principal directions of PCA from LAE weight matrices. This is as follows:
\begin{itemize}
    \item train a $L_2$-regularized LAE with loss function $\mathcal{L}_\sigma$ (input data can be not mean-scaled). The optimal $W_1$ and $W_2$ are $W_1^*$ and $W_2^*$;
    \item apply SVD on $W_2^{*T}$ ($T \times K$): $W_2^{*T} = U \Sigma V^T$;
    \item the loading vectors are the columns of $U$ i.e. the left singular
vectors of the decoder.
\end{itemize}
This algorithm is a consequence of the \textit{Landscape Theorem} of the paper \cite{kunin2019}. Indeed, according to this Theorem, we have that the optimal value of the decoder and the encoder matrices for $\mathcal{L_\sigma}$ are defined up to an orthogonal map $O \in O_K(\mathbb{R})$:
\begin{equation*}
    W_2^T = U_K (I - \lambda \Sigma_K^{-2})^{\frac{1}{2}}O = W_1
\end{equation*}
where $X = U \Sigma V^T$ and $\sigma_1^2 > \sigma_2^2 > \ldots > \sigma_K^2 > \lambda$. In the last equality, the \textit{Transpose Theorem} \cite{kunin2019} has been employed: it states that all critical points of $\mathcal{L}_\sigma$ satisfy $W_1 = W_2^T$.

To sum up, the $L_2$-regularized LAEs are
transposes at all critical points and learn the principal directions as the left singular vectors of the decoder. Given this relation between LAE and PCA and the algorithm above, using LAE instead of PCA could be useful for large datasets. Indeed, SVD will be performed on a smaller matrix $W_2^*$ which is $K \times T$, instead of $X$ that is $N \times T$. Moreover, having a PCA-like solution allows to exploit nested solutions easily. Indeed, if results are obtained for a given $K$ then, we can obtain the solution for $K'\neq K$, by truncating the loading vectors' matrix $U$ at $K'$ instead of $K$.

Finally, we recall that as proved in \cite{zhou2018}, the loss function for linear networks has no spurious local minimum, while such point does exist for
nonlinear networks with ReLU activation.

\subsubsection{Results}
\begin{figure}[!t]
    \centering\includegraphics[width=.45\linewidth]{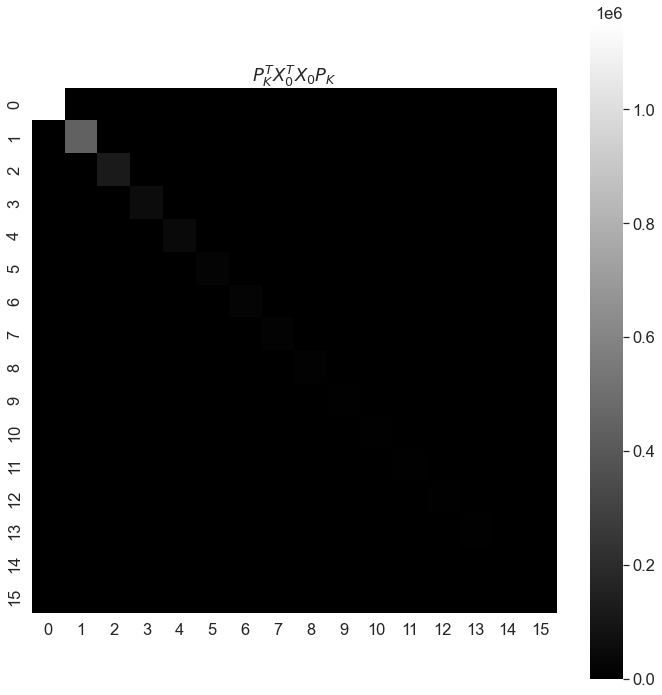}
    \caption{Covariance matrix of $X_0P_K$ i.e. PCA compressed representation of the mean-centered data.}
    \label{cov_matrix_XP}
\end{figure}

\begin{figure}[!t]
    \centering\includegraphics[width=.45\linewidth]{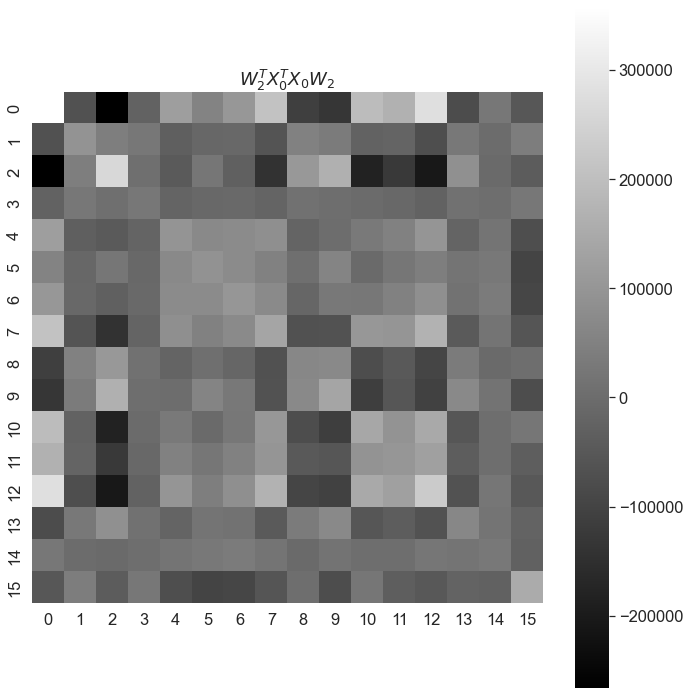}
    \caption{Covariance matrix of $X_0W_1$ i.e. LAE compressed representation of the mean-centered data.}
    \label{cov_matrix_XW}
\end{figure}

\begin{figure}[!t]
    \centering\includegraphics[width=.45\linewidth]{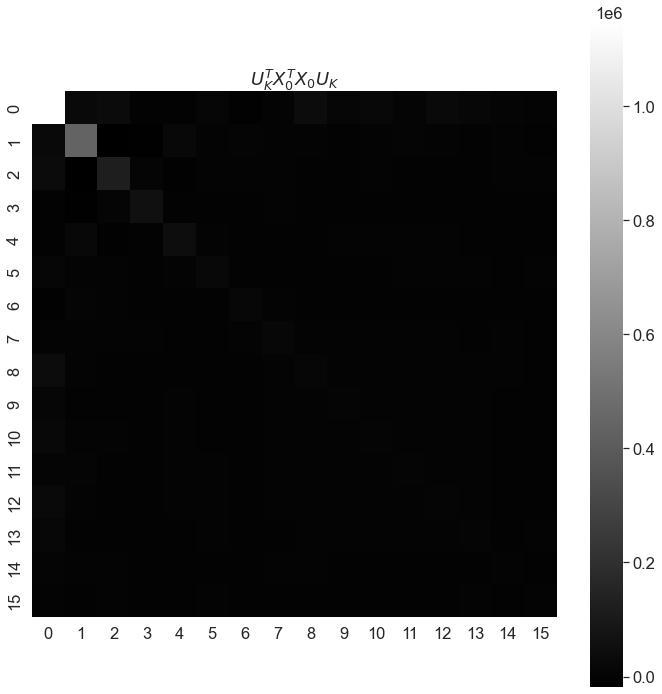}
    \caption{Covariance matrix of $X_0U_K$ i.e. compressed representation of the mean-centered data, using the loading vectors obtained by the LAE.}
    \label{cov_matrix_XU}
\end{figure}

\noindent Let us consider the case study related to IMA using  $K=16$ and performing the dimensionality reduction step with a LAE. We would like to test the analogy between LAE and PCA, by relying on the results of \cite{kunin2019} and as explained in the previous paragraph. Therefore, our architecture is a $L_2$-regularized LAE with one hidden layer with $K$ neurons. We apply three different transformations to the mean-centered data $X_0$ i.e. $X_0P_K$, $X_0W_1$, $X_0U_K$ where $P_K$ is the loading vectors' matrix obtained by PCA, $W_1$ is the encoder of the LAE and $U_K$ is the loading vectors' matrix obtained by the LAE. In Figures \ref{cov_matrix_XP}-\ref{cov_matrix_XU}, the covariance matrices of these transformed data are represented. As expected according to \cite{kunin2019}, the covariance matrix is diagonal and with descending diagonal elements for $X_0P_K$ and $X_0U_K$; this is not the case for the covariance matrix of $X_0W_1$.

\section{Anomaly detection with autoencoders}\label{ad_ae}
\noindent In this appendix, figures concerning the results related to our method based on the employment of autoencoders and applied on the asset IMA, are provided. Explanations of these results are inserted in the main text.

\begin{figure}[!t]
    \centering\includegraphics[width=.45\linewidth]{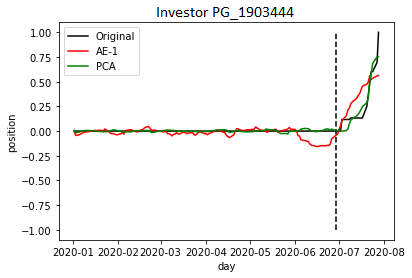}
    \caption{IMA. Position of an investor detected by the method based on AE-1 and not by the
method based on PCA. The vertical dashed line is the day corresponding to the beginning of the investigation period i.e. June 29, 2020.}
\label{PG_1903444_AE_1}
\end{figure}


\begin{figure}[!t]
    \includegraphics[width=.415\linewidth]{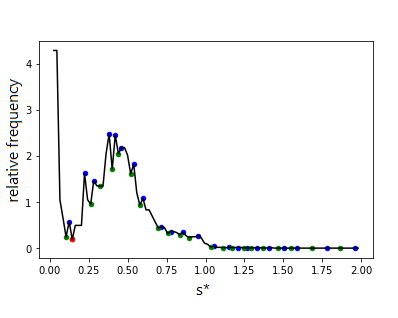}    \includegraphics[width=.45\linewidth]{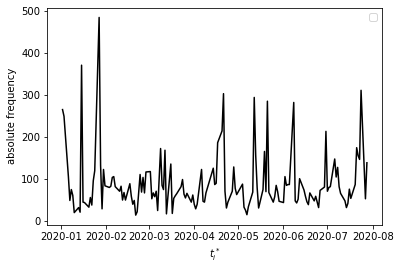}
    \label{IMA_anomaly_score_pdf_AE_4.png}
    \caption{AE-4 on IMA. Left. Histogram of the anomaly scores; blue points are local maxima, green local minima and the
red point is $\epsilon_\theta \simeq 0.14$ that is the local minimum after the first peak. Right. Histogram of the times corresponding to the anomaly scores.}
\end{figure}

\section{Households versus firms}
\noindent As we explained in Section \ref{section_data}, the data set we are provided with, contains information about each investor type. We consider two broad categories of investors: households, that include individual households and joint accounts of several households, and firms, that include investment firms and other legal entities. The goal of this appendix is to investigate whether it could be more advantageous to run our whole methodology separately for households and firms. This is motivated by the fact that, in principle, these two classes of investors have different behavior. 

\begin{table*}
\begin{center}
    \caption{IMA and UBI. Composition of the two groups of investors in the latent space and in the anomaly score distribution.}
    \label{table_households_vs_firms}
    \begin{tabular}{c|c|c||c|c}
    & \multicolumn{2}{c||}{IMA} & \multicolumn{2}{c}{UBI} \\
    \hline
    Investors' type &  Households & Firms & Households & Firms \\
    \hline
         Group 1 latent space & $9,734$  & $453$ & $12,829$ & $306$\\
         \hline
         Group 2 latent space& $2,846$ & $192$ & $18,027$ & $808$ \\
         \hline
         \hline
        A: Investors with $s_i^* \geq \epsilon_\theta$ &$11,011$ & $606$ & $26,789$ & $686$\\
        \hline
        B: Investors with $s_i^* < \epsilon_\theta$ & $1,569$  & $39$ & $4,067$ & $428$\\
    \end{tabular}
    \end{center}
\end{table*}

\subsection{PCA using all data: household-firm composition}
\noindent Let us consider our case study related to the asset IMA. We have $12,580$ households and $645$ firms. As expected, the dataset is highly imbalanced towards households who constitute $95.1$\% of investors. However, their corresponding exchanged volume is less than $10$\% of the total. 

We consider the representation in the latent space obtained by PCA which uses all data. A clustering method (the k-means) for two groups is run and we check whether one of the group is mainly composed of households and the other one of firms. The groups' composition is reported in Table \ref{table_households_vs_firms}. We perform a Fisher test with the null hypothesis that there is not association between groups in the latent space and investor types. The p-value turns out to be $4.3$e-$5$ so, we need to reject the null hypothesis: there is a relation between investor type and group. Analogously, we test whether a given investor type is over/under-expressed in a group, as in \cite{svn1}. For this test, the null hypothesis is defined by assuming the random co-occurrence of a given investor type and her belonging to a given group. The hypergeometric distribution is used as a benchmark for randomness. It results that in group 1 (2), the investor type \emph{household} (\emph{firm}) is over-expressed and the investor type \emph{firm} (\emph{household}) is under-expressed.

However, given that our method relies on the computation of the reconstruction errors, we further investigate the household-firm composition in the anomaly score distribution obtained by running PCA using the whole dataset. We split investors in two categories: investors with anomaly score $s_i^*$ greater than or equal to the threshold $\epsilon_\theta$ (group A) and investors with anomaly score lower than the threshold (group B). The groups' composition is shown in Table \ref{table_households_vs_firms}. Also in this case, the Fisher test points out that there is association between the investor type and the group in the anomaly score distribution. We test the over/under-expression of investor types in the two groups, as in \cite{svn1}. It results that in group A (B), the investor type \emph{firm} (\emph{household}) is over-expressed and the investor type \emph{household} (\emph{firm}) is under-expressed. We can conclude that, basically, firms are associated with higher values of anomaly score. 

The results reported so far are related to the asset IMA, which is illiquid and, as shown in \cite{our_paper}, exhibits strong synchronization signals related to the PSE under investigation. If we consider the asset UBI, which is much more liquid than IMA, the findings related to the latent space representation, are analogous. On the other hand, if we focus on the composition household-firm in the anomaly score distribution, we find that in group A (B), the investor type \emph{household} (\emph{firm}) is over-expressed and the investor type \emph{firm} (\emph{household}) is under-expressed. Therefore, contrary to IMA, higher scores are associated with households. This difference between IMA and UBI could be explained by the fact that, as mentioned above, in \cite{our_paper} we showed that investors trading IMA were having strong synchronization signals related to the PSE. Indeed, in the second clustering approach of \cite{our_paper}, based on the statistically validated co-occurrence networks and aimed at identifying groups of investors with coordinated suspicious behavior related to the PSE, we identify an highly synchronized cluster made up of more than $2,000$ investors, who are mainly households and with the portfolios managed by the same entity. This issue together with the fact that IMA's data set is small, could have make easier reconstructing households' profiles.

\subsection{PCA using households' and firms' data separately}\label{subsection_PCA_h_vs_F}
\noindent We perform PCA using the datasets made up of the two categories of investors separately. We compare the results between them and with the results obtained by running PCA with the whole dataset. 

\begin{figure}[!t]
    \includegraphics[width=.45\linewidth]{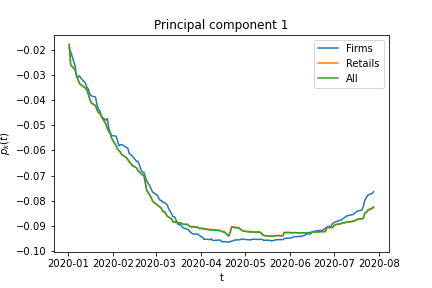}
    \includegraphics[width=.45\linewidth]{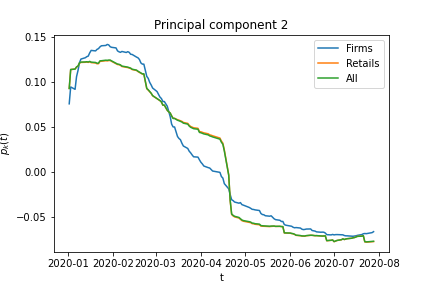}
    \includegraphics[width=.45\linewidth]{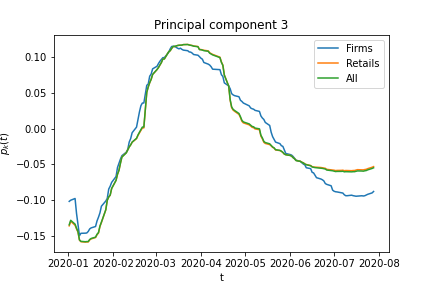}
    \includegraphics[width=.45\linewidth]{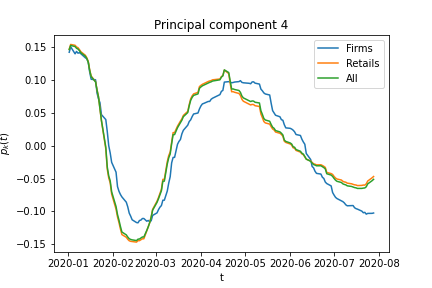}
    \includegraphics[width=.45\linewidth]{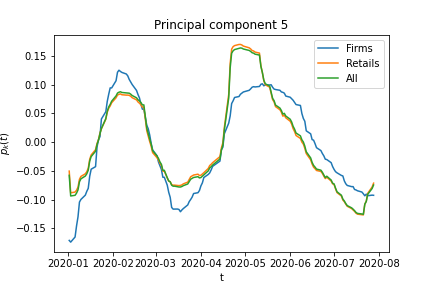}
    \includegraphics[width=.45\linewidth]{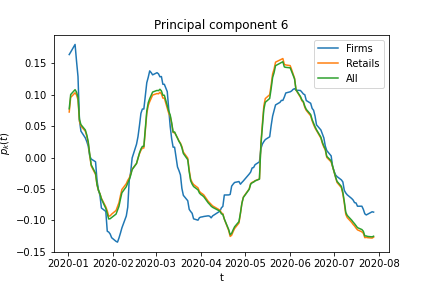}
    \centering\includegraphics[width=.45\linewidth]{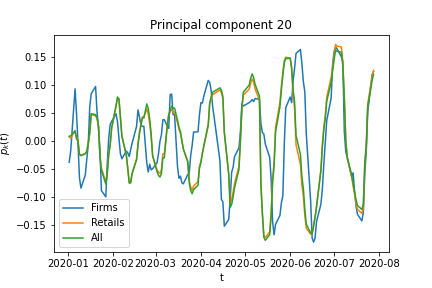}
    \caption{IMA. Representation of some of the principal components obtained by performing PCA using the whole dataset (\emph{All}), the dataset made up of households (\emph{Retails}) and the dataset made up of firms (\emph{Firms}).}
\label{principal_components_households_vs_firms}
\end{figure}

\begin{figure}[!t]
    \centering\includegraphics[width=.45\linewidth]{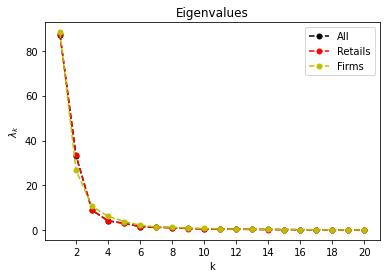}
    \caption{IMA. Eigenvalues obtained by performing PCA using the whole dataset (\emph{All}), the dataset made up of households (\emph{Retails}) and the dataset made up of firms (\emph{Firms}).}
\label{eigenvalues_households_vs_firms}
\end{figure}

Let us start to focus on IMA. Given the extremely high fraction of households ($95.1 \% $), the difference between PCA results obtained by considering only households and by considering all the investors is negligible. The differences between PCA results obtained by considering only households or only firms are not substantial, especially for the first components, i.e. the components which retain more data variability. This is shown in Figure \ref{principal_components_households_vs_firms} where the first $6$ components and the twentieth component are shown (let us recall that each principal component is a vector of dimensionality $T$). Moreover, in Figure \ref{eigenvalues_households_vs_firms}, a comparison between the eigenvalues obtained is provided. Analogous results are obtained for UBI.

\subsection{Insider trading detection using households' and firms' data separately}
\begin{figure}[!t]
    \centering
    \includegraphics[width=.45\linewidth]{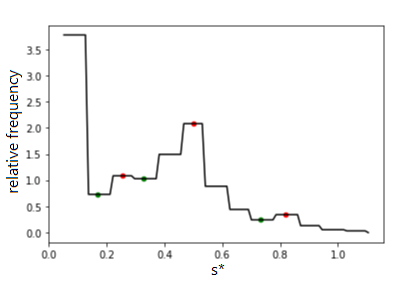}
    \caption{UBI. Histogram of the anomaly scores obtained by using PCA with $K=16$; red points are local maxima, green local minima.}
    \label{anomaly_score_distribution_ubi_only_firms}
\end{figure}

\noindent Now, let us tackle our major goal of this appendix, that is investigating whether it could be more advantageous to run our whole methodology for insider trading detection, separately for households and firms. As we illustrated in the previous paragraph, for IMA (UBI), firms (households) are associated with higher values of anomaly score and households (firms) with lower values of anomaly score. For IMA, this issue together with the small number of firms ($606 + 39$) imply that, if we perform PCA using only the data related to firms, the anomaly score distribution we obtain, does not show the bimodality we want to exploit in order to set the threshold $\epsilon_\theta$, which has a major role in the criterion of Equation \ref{eq_anom}. On the other hand, for UBI, the higher number of firms (in absolute value) allows to preserve the bimodality of the anomaly score distribution obtained by running PCA with only the data related to firms, as shown in Figure \ref{anomaly_score_distribution_ubi_only_firms}. Therefore, we focus on UBI for the subsequent analysis. 

\begin{figure}[!t]
    \centering
    \includegraphics[width=.45\linewidth]{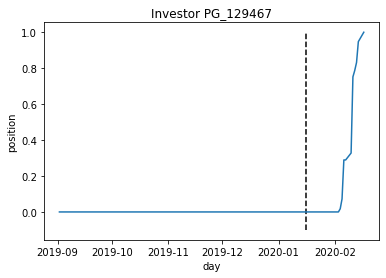}
    \caption{UBI. Profile of an investor identified as anomalous by using only the dataset made up of firms and not by using the whole dataset. The dotted black line is the beginning of the investigation period.}
    \label{PG_129467_UBI_only_firms}
\end{figure}

We apply our whole methodology to identify potential insiders, for households and firms separately: results are shown in Table \ref{tab_ubi_households_vs_firms_anomalous}. The method which uses only the data related to firms identifies $16$ more anomalous investors than the method which uses the whole dataset. Some of them, like the profile in Figure \ref{PG_129467_UBI_only_firms}, could be interesting for our scope. However, if we focus on the first $500$ ranked potential insiders, there is no difference. On the other hand, if the methodology is run by using only the data related to households, a consistent number of potential insiders is not identified with respect to the method which uses all the data and, among the first $500$ ranked potential insiders, $59$ investors are not detected. These households are actually extremely suspicious since they are all just active in the investigation period with a net buying position, similarly to the profile in Figure \ref{PG_129467_UBI_only_firms}.

The difference between the results obtained by using all data and the data only related to households, could be surprising: in Subsection \ref{subsection_PCA_h_vs_F}, we observed that the difference between the principal components and the eigenvalues obtained in the two cases is negligible. However, it is important to remember that in the criterion of Equation \ref{eq_anom}, also the times corresponding to the largest reconstruction errors $t_i^*$ have a role and in fact, using the data only related to households, causes a change in the $t_i^*$ histogram.

\begin{table*}
\begin{center}
    \caption{UBI. $A_c$ with $c=\{\text{households}, \ \text{firms}\}$ is the set of potential insiders obtained by using only the data related to investors of type $c$. $A^{all}$ is the set of potential insiders obtained by using all data. $C$ is the set of households/firms in the data set. $A^{all}_{500}$ and $A_{500}^c$ are the set of the first $500$ ranked potential insiders obtained by using all data or only the data of investors of type $c$ respectively.}
    \label{tab_ubi_households_vs_firms_anomalous}
    \begin{tabular}{c|c|c|c|c|c|c}
    Investors' type & $|A^c|$ & $|A^{all} \cap C|$ & $|A^c \cap A^{all}|$ & $|A^{all}_{500} \cap C|$ & $|A^c \cap A^{all}_{500}| $  & $|A^c_{500} \cap A^{all}_{500}|$ \\
    \hline
       Households  & $1,580$ & $1,722$ & $1,580$ & $491$ & $483$ &  $432$\\
         Firms & $95$ & $79$ & $78$ & $9$ & $9$ & $9$
    \end{tabular}
    \end{center}
\end{table*}

To conclude, we verified that if PCA using all data is performed, there is a split between households and firms, both in the latent space representation and in the anomaly score distribution. However, for small assets as IMA, the anomaly score distribution loses its bimodality once PCA is applied using the data related to only firms. Thus, setting the threshold to run our \textit{reconstruction-based} approach, is problematic. This does not occur for more liquid assets as UBI, for which the number of firms, even if it is less than $5\%$ of the total number of investors, is greater. We find that for UBI, performing our method using the data related to the two investors' classes separately leads to an improvement for firms if we go beyond the first $500$ ranked anomalous investors. On the other hand, for households, it leads to a deterioration in our results. Investors who have a net suspicious activity related to the PSE, are missed. Therefore, running our whole methodology separately for households and firms does not seem to be consistently more advantageous.

\end{document}